\pdfoutput=1
\documentclass[letterpaper,11pt]{article}
\usepackage{amsmath,amssymb,amsfonts}
\usepackage[multiple,stable]{footmisc}
\allowdisplaybreaks[4]
\usepackage[noconfig]{refstyle}
\usepackage[dvipsnames]{xcolor}
\usepackage[hyperfootnotes=false, linktocpage=true, colorlinks, citecolor=blue, linkcolor=blue, urlcolor=Maroon]{hyperref}
\usepackage{array,tabularx,booktabs,subfigure,graphicx,longtable}
\usepackage[bf]{caption}
\usepackage[paper=letterpaper,margin=1in]{geometry}
\usepackage{tikz}
\usetikzlibrary{shapes,matrix,arrows}
\usepackage{url,lineno}
\usepackage{bm}
\usepackage{cancel}
\usepackage{enumitem}
\usepackage{listings}
\usepackage{verbatim}
\usepackage{tensor}
\usepackage{bbm}
\usepackage{mathtools}
\usepackage{graphicx}
\usepackage{listings}
\usepackage{caption}
\usepackage{tikz}
\usetikzlibrary{trees}
\usepackage{cite}
\usepackage{blkarray}
\usepackage{float}

\usepackage{listings}
\usepackage{hyperref}
\usepackage{xcolor}

\definecolor{bgd}{rgb}{.88,.87,.88}
\definecolor{sage}{rgb}{.35,.21,.85}
\usepackage{appendix}

\usepackage{indentfirst}

\tikzstyle{block} = [rectangle, draw, text width=7em, text centered, rounded corners, minimum height=3em]

\def\tcA{\tilde{\mathcal{A}}}
\newcommand{\matindex}[1]{\mbox{\scriptsize#1}}
\newcommand{\LD}[1]{[#1]_{lin}}
\newcommand{\AND}[1]{\hspace{5mm}\text{#1}\hspace{5mm}}
\newcommand{\RD}[1]{[#1]_{rat}}
\newcommand{\ND}[1]{[#1]_{num}}
\newcommand{\cA}{{\cal A}}
\newcommand{\iddots}{\mathinner{\mkern2mu\raise1pt\hbox{.}\mkern2mu \raise4pt\hbox{.}\mkern2mu\raise7pt\hbox{.}\mkern1mu}}

\providecommand{\id}{\leavevmode\hbox{\small$\mathrm{1}$\kern-3.8pt\normalsize$\mathrm{1}$}}
\def\fnote#1#2{\begingroup\def\thefootnote{#1}\footnote{#2}
     \addtocounter{footnote}{-1}\endgroup}

\begin{document}

\vspace{1cm}

\title{\begin{flushright}
       \mbox{\footnotesize WITS-CTP-151}
       \end{flushright}
       \vskip 40pt
\textbf{      {\Huge A Calabi--Yau Database}\\
Threefolds Constructed from the Kreuzer--Skarke List}}
      \vspace{2cm}

\author{\textbf{Ross Altman${}^a$, James Gray${}^b$, Yang-Hui He${}^{c,d,e}$, Vishnu Jejjala${}^f$, Brent D.\ Nelson${}^{a,g}$}}
\date{}
\maketitle
\begin{center}${}^a${\small  Department of Physics, Northeastern University, Boston, MA 02115, USA} \\
${}^b${\small Physics Department, Robeson Hall, Virginia Tech, Blacksburg, VA 24061, USA} \\
${}^c${\small Department of Mathematics, City University, London, EC1V 0HB, UK} \\
${}^d${\small School of Physics, NanKai University, Tianjin, 300071, P.R.\ China} \\
${}^e${\small Merton College, University of Oxford, OX1 4JD, UK}\\
${}^f${\small Centre for Theoretical Physics, NITheP, and School of Physics, \\
University of the Witwatersrand, Johannesburg, WITS 2050, South Africa} \\
${}^g${\small ICTP, Strada Costiera 11, Trieste 34014, Italy}\\
\fnote{}{${}$\hspace{-0.25in}altman.ro@husky.neu.edu, grayphys@vt.edu, hey@maths.ox.ac.uk, vishnu@neo.phys.wits.ac.za, b.nelson@neu.edu}
\end{center}

\begin{abstract}
Kreuzer and Skarke famously produced the largest known database of Calabi--Yau threefolds by providing a complete construction of all 473,800,776 reflexive polyhedra that exist in four dimensions~\cite{Kreuzer:2000xy}.
These polyhedra describe the singular limits of ambient toric varieties in which Calabi–-Yau threefolds can exist as hypersurfaces.
In this paper, we review how to extract topological and geometric information about Calabi--Yau threefolds using the toric construction, and we provide, in a companion online database (see \url{http://nuweb1.neu.edu/cydatabase}), a detailed inventory of these quantities which are of interest to physicists.
Many of the singular ambient spaces described by the Kreuzer--Skarke list can be smoothed out into multiple distinct toric ambient spaces describing different Calabi--Yau threefolds.
We provide a list of the different Calabi--Yau threefolds which can be obtained from each polytope, up to current computational limits.
We then give the details of a variety of quantities associated to each of these Calabi--Yau such as Chern classes, intersection numbers, and the K\"ahler and Mori cones, in addition to the Hodge data.
This data forms a useful starting point for a number of physical applications of the Kreuzer--Skarke list.
\end{abstract}

\thispagestyle{empty}
\setcounter{page}{0}
\newpage

\setcounter{tocdepth}{2}
\tableofcontents

\numberwithin{equation}{section}
\def\theequation{\thesection.\arabic{equation}}

\renewcommand\tilde{\widetilde}

\section{Introduction}\label{intro}

Calabi--Yau manifolds provide one of the simplest ways to compactify the extra dimensions of string theory while preserving some supersymmetry in four dimensions.
Following the original idea of Kaluza and Klein~\cite{Kaluza:1921tu,Klein:1926tv}, the topology and geometry of the extra dimensions determines the effective dynamics seen in the lower dimensional theory.
Such geometrical constructions offer a rich framework for string phenomenology and more formal studies in a variety of different string theoretic contexts.

A compelling example of the use of Calabi--Yau threefolds in string theory can be seen in the subject of heterotic string phenomenology, as originally described in~\cite{Candelas:1985en}.\footnote{
It should be emphasized that there is little consensus within the field as to which string theoretic construction is most likely to lead to realistic particle physics, with each approach having its own strengths and weaknesses.}
During the last decade, semi-realistic constructions exhibiting precisely the charged matter spectrum of the Standard Model of particle physics have been derived many times within this context~\cite{Braun:2005ux,Braun:2005bw,Bouchard:2005ag,Braun:2005nv,Anderson:2009mh,Anderson:2012yf}.
The key point for this paper is that, given the link between geometry and low energy physics mentioned above, in all such discussions knowledge of the geometry of the Calabi--Yau threefolds involved is crucial in making progress.

Over the last three decades various large datasets of explicit constructions of Calabi--Yau manifolds have been developed, ranging from the so-called complete-intersection Calabi--Yau manifolds (CICYs)~\cite{Hubsch:1986ny,Candelas:1987kf,Green:1986ck,Candelas:1987du,Brunner:1996bu,Gray:2013mja,Gray:2014fla} to elliptically fibered manifolds over toric bases~\cite{Morrison:2012js,Anderson:2014gla} and hypersurfaces in toric ambient spaces~\cite{Kreuzer:2000qv,Kreuzer:2000xy,Berglund:1991pp}.
By far the largest set found to date is given by the impressive work due to Kreuzer and Skarke in the 1990s: realizing Calabi--Yau threefolds as a hypersurfaces in four-dimensional toric varieties~\cite{Kreuzer:2000qv,Kreuzer:2000xy}. A huge variety of different data sets have been isolated from this initial work. While all of the different applications are too numerous to list here, some interesting examples of such work can be found in these references~\cite{Blumenhagen:2012kz,Gao:2013pra,Gao:2013rra,Cicoli:2012vw,Cicoli:2011it,Cicoli:2008va,Cicoli:2013cha,Batyrev:2005jc}.

Batyrev showed that a hypersurface in a toric variety can be chosen to be Calabi--Yau if the object underlying the construction of the variety, a lattice polytope, obeys the condition of reflexivity~\cite{Batyrev:1994hm}.
The classification of inequivalent reflexive polytopes is an exercise in combinatorics and is thus amenable to computer analysis.
The software {\tt PALP} (Package for Analyzing Lattice Polytopes)~\cite{Kreuzer:2002uu} was written with precisely this goal in mind.
In all, an impressive 473,800,776 four-dimensional reflexive polytopes were found, each of which can be resolved to give candidate ambient spaces for Calabi--Yau threefold hypersurfaces.\footnote{
For context, one might compare this to the case of two- and three-dimensional toric ambient spaces giving rise to Calabi--Yau one- and two-fold hypersurfaces (i.e., the torus and K3 surfaces).
In these dimensions there are only 16 and 4,319 polytopes, respectively.}
Some important topological information such as the Hodge numbers for the Calabi--Yau associated to a given polytope can be found in the extremely useful website~\cite{Kreuzera}.
Here, and in the accompanying website~\cite{Altmana}, we wish to add to what can be found in~\cite{Kreuzera}, in order to fill a number of gaps in the information that is currently publicly available.

\begin{enumerate}
\item
Although the number of distinct reflexive polytopes in the Kreuzer--Skarke database is well understood, it is unclear precisely how many distinct Calabi--Yau threefolds emerge from this list.
Some of the geometrical data of these manifolds, such as Hodge numbers, can be determined purely in terms of the ambient polytope data.
Of the 473,800,776 reflexive polytopes, it has been shown that there are 30,108 distinct pairs $(h^{1,1},h^{2,1})$ counting the K\"ahler and complex structure moduli of the geometries respectively.
However, different so-called triangulations into simplexes of the same polytope can potentially give rise to different Calabi--Yau hypersurfaces, which, while agreeing on this basic geometric data, differ in more subtle ways.
In short, even taking into account the redundancy present in describing the same Calabi--Yau manifolds as hypersurfaces in different ambient spaces, there are likely to be many more than 473,800,776 Calabi--Yau threefolds in the Kreuzer--Skarke list.
Unfortunately, performing all of the necessary triangulations to access this data is computationally intensive, and {\tt PALP} is only able to complete the necessary computations for relatively low values of $h^{1,1}$~\cite{Braun:2011ik}.
Clearly, therefore, it is of use to have as much of this data as possible pre-computed and archived in an easy-to-access format.

\item
For calculations in subjects such as model building (for example in the heterotic case~\cite{He:2009wi,He:2011rs,He:2013ofa}), more than just the Hodge numbers of a Calabi--Yau threefold are required.
The Chern classes, K\"ahler cones, and  triple-intersection numbers, which help to characterize the Calabi--Yau threefold in a more refined manner, are also needed, for example.
These quantities can again  be extracted from various versions, some not yet publicly available, of {\tt PALP}.
As in the previous bullet point, it is clear that we would like to have as much of this data as possible pre-computed and archived in an easy to access format.

\item
A variety of codes are available for calculating geometrical properties of Calabi--Yau threefold hypersurfaces in toric varieties.
Building upon the original work of Max Kreuzer and his collaborators, there have been some updated versions of \texttt{PALP}, consolidating and enhancing his contributions while providing additional documentation for the program~\cite{Braun:2011ik,Braun:2012vh}.
In addition, new tools for analyzing such geometries are now available within the context of {\tt Sage}~\cite{sage} and~{\tt TOPCOM}~\cite{Braun2011}.
In addition to being computationally intensive to run over large data sets, the language of programs such as {\tt PALP} assume a level of detailed mathematical knowledge which, for physicists who are unfamiliar with the subject and simply wish to extract certain properties of the Calabi--Yau manifolds, may seem somewhat onerous to learn.
Once more we see that it would be useful to have the data that physicists are interested in pre-computed and archived in a physicist's language.
\end{enumerate}

Given the above, our purpose in writing this paper is twofold.
Firstly, we review the computational procedures for calculating the relevant topological and geometrical quantities of toric Calabi--Yau threefolds in a manner that is completely self-contained within the {\tt Sage} computational package.
Second, this paper supplies a user's manual for how to quickly extract the Calabi--Yau data from the webpage~\cite{Altmana}, which provides an archive of this data in a pre-computed and easy-to-access format.
Because of limitations in the computational power that has been applied to the problem to date, the database associated to this paper will provide all systematic triangulations up to $h^{1,1}=6$ and provide partial results for $h^{1,1}=7$.
This already exceeds what be can be accessed with {\tt PALP}, and importantly, the physicist accessing this data need not run or become familiar with any additional software.
We expect to update the website as time goes by to accommodate ever increasing $h^{1,1}$.

The organization of this paper is as follows.
For readers who are familiar with algebraic geometry but who have not studied toric varieties, we aim, in Section~\ref{sec:pedagogy}, to provide a pedagogical explanation of how practically to obtain geometrical data describing a Calabi--Yau threefold starting from an element of the dataset of~\cite{Kreuzer:2000qv,Kreuzer:2000xy}.
Readers familiar with this subject should skip directly to Section~\ref{sec:database}, where we explain, with examples, how to extract a wide variety of information from the database of Calabi--Yau threefolds that we have derived from the Kreuzer--Skarke classification.
Finally, we conclude with a discussion in Section~\ref{sec:conc}.
For the reader who is familiar with differential geometry but is not so comfortable with algebraic concepts, we provide in Appendix~\ref{appa} a basic introduction to some of the notions which are used in the text.
Finally, in Appendix~\ref{glossary}, we provide a glossary of the nomenclature used throughout this paper.


\section{Methods: Calabi--Yau Threefolds from Toric Construction}
\label{sec:pedagogy}

In this section, we will review how to extract relevant topological and geometrical information about Calabi--Yau threefolds, starting from reflexive polytopes.
Due to the work of Kreuzer and Skarke~\cite{Kreuzera,Kreuzer:2000xy}, not only do we know that the set of four-dimensional reflexive polytopes is finite, but we actually have a complete database of them to draw on.
In the subsequent section we will present the results of using the methods we discuss here to convert the database of four-dimensional reflexive polytopes \cite{Kreuzera} to a database of Calabi--Yau threefold properties \cite{Altmana}.
For a brief introduction to some of the key algebro-geometric concepts used in this section please refer to Appendix~\ref{appa}.

In what follows, we will assume the ambient space $\cA$ to be a Gorenstein toric Fano variety with dimension $n=4$ (although we will sometimes present results in general $n$) whose anticanonical divisor $X=-K_{\mathcal{A}}$ is a Calabi--Yau threefold hypersurface.
As such, the Newton polytope $\Delta$, corresponding one-to-one with $\cA$ is a reflexive polytope, which implies that $\Delta$ as well as its dual $\Delta^{*}$ are lattice polytopes containing only the origin in their respective interiors.
Also, for simplicity of notation, we will represent all linear equivalence classes of divisors $\LD{D}$ by a representative divisor $D$.
A more comprehensive explanation of our notation is provided in Appendix~\ref{glossary}.

\subsection{Kreuzer--Skarke Database}

Kreuzer and Skarke created an algorithm to generate all reflexive polytopes in dimension $n\geq 4$.
Fortunately for us, we are only interested in dimension $n=4$.
It turns out that there are 473,800,776 reflexive polytopes with $n=4$.

The output of the Kreuzer--Skarke (KS) database~\cite{Kreuzera} is a text file with every 5 (i.e., $n+1$) lines corresponding to a reflexive polytope.
The first line is a summary of certain key information about the Newton polytope $\Delta$, its dual $\Delta^{*}$, and the toric variety $\cA$ they encode.
It reads:\\

\noindent \texttt{<$\textrm{dim}(\Delta)$> <$\textrm{card}(\mathcal{V}(\Delta))$>\\  M:<$\textrm{card}(\Delta)$> <$\textrm{card}(\mathcal{V}(\Delta))$>\\ N:<$\textrm{card}(\Delta^{*})$> <$\textrm{card}(\mathcal{V}(\Delta^{*}))$>\\ H:<$h^{1,1}(X)$,$h^{2,1}(X)$> [<$\chi(X)$>]}\\

\noindent Recall from Appendix~\ref{glossary} that the notation $\mathcal{V}(\Delta)$ represents the set of vertices of the reflexive polytope $\Delta$, and $\text{card}(P)$ indicates the cardinality, or number of lattice points, in any subspace $P\subset M$ or $N$.

The remaining four ($n$) lines contain a matrix whose columns are the vertices $\bm{m}\in\mathcal{V}(\Delta)$ of the Newton polytope $\Delta$.
In this section, we will occasionally refer to the $95^{\text{th}}$ polytope with $h^{1,1}(X)=3$, $\Delta_{3,95}$ as an example.
It is given in the KS database as:\\
\\
\begingroup
\tt
4 6  M:100 6 N:8 6 H:3,81 [-156]\\
\begin{tabular}{rr r r r r r}
& 1 & 1 & 1 & 1 & -5 & -5\\
& 0 & 3 & 0 & 0 & -6 & 0\\
& 0 & 0 & 3 & 0 & 0 & -6\\
& 0 & 0 & 0 & 3 & -3 & -3
\end{tabular}
\endgroup

\subsection{Parsing the Database using \texttt{Sage}}

We first extract the necessary information from the database entry.
The database directly gives us the vertices $\mathcal{V}(\Delta)$ of the Newton polytope.
It also gives us the Hodge numbers ($h^{1,1}(X), h^{2,1}(X)$) and the Euler number $\chi(X)$ of a Calabi--Yau hypersurface $X\subset\cA$, but we will recalculate these later for the sake of completeness.

{\tt Sage} allows us to define the polytope $\Delta$ from vertices $\mathcal{V}(\Delta)$ using the \texttt{Polyhedron} class.
The dual (or polar) polytope $\Delta^{*}$ can then be obtained using the \texttt{Polyhedron$\rightarrow$polar()} method.
From these, we can easily obtain the lattice points of $\Delta$ and $\Delta^{*}$ with \texttt{Polyhedron$\rightarrow$integral\_points()} as well as the vertices $\mathcal{V}(\Delta^{*})$ with \texttt{Polyhedron$\rightarrow$vertices()}.

We then obtain the faces $\mathcal{F}(\Delta)$ and $\mathcal{F}(\Delta^{*})$ using the H(ypersurface) representation \newline\texttt{Polyhedron$\rightarrow$Hrepresentation()}.
From this, it is a simple matter to construct the cones $\sigma_{F}=\text{cone}(F)$ using the \texttt{Cone} class.
The cones of $\Delta$ and $\Delta^{*}$ can each be joined into fans $\Sigma(\Delta)$ and $\Sigma(\Delta^{*})$ using the \texttt{Fan} class.

\subsection{Stringy Hodge Numbers and Euler Number}\label{sec:hodge}

The stringy Hodge numbers for a generic Calabi--Yau hypersurface $X\subset\cA$ can be computed using Batyrev's well-known formulae \cite{Batyrev:1994hm} :

{\small
\begin{align}
h^{1,1}(X)&=\text{card}(\Delta^{*})-5-\!\!\!\!\!\!\sum_{\text{codim}(F^{*})=1}\!\!\!\!\!\!{\text{card}(\text{relint}(F^{*}))}+\!\!\!\!\!\!\sum_{\text{codim}(F^{*})=2}\!\!\!\!\!\!{\text{card}(\text{relint}(F^{*}))\cdot\text{card}(\text{relint}(F))}\notag\\
&\notag\\
h^{2,1}(X)&=\text{card}(\Delta)-5-\!\!\!\!\!\!\sum_{\text{codim}(F)=1}\!\!\!\!\!\!{\text{card}(\text{relint}(F))}+\!\!\!\!\!\!\sum_{\text{codim}(F)=2}\!\!\!\!\!\!{\text{card}(\text{relint}(F))\cdot\text{card}(\text{relint}(F^{*}))}\, .
\end{align}
}
The Euler number is then easily computed via
\begin{equation}
\chi(X)=2\cdot (h^{1,1}(X)-h^{2,1}(X))\, .
\end{equation}

\subsection{MPCP Desingularization and Triangulation of $\Delta^{*}$}\label{sec:MPCP}

The variety $\cA$ generated by $\Delta$ may be a singular space.
If it is too singular, then our Calabi--Yau hypersurface $X\subset\cA$ may not be smooth even though it is base point free.
Therefore, we must find an appropriate resolution of singularities given by a birational morphism $\pi :\;\tcA\rightarrow\cA$ such that the desingularized space $\tcA$ is smooth enough that the hypersurface $X\subset\tcA$ can be chosen smooth.
Because $X$ has dimension 3, it can generically be transversally (i.e., smoothly) deformed around singular loci with codimension 3.
Such singular loci with codimension $\geq 3$ are called \textit{terminal singularities} \cite{Matsuki2002}.
Thus, we need only consider partial desingularizations which resolve everything up to terminal singularities.

Another important condition for $X$ to be smooth is for it to be well-defined everywhere when viewed as a Cartier divisor.
For this to be true, we must be able to write $X$ uniquely in terms of a basis on every coordinate patch $U$ on the open cover of $\tcA$.
Because an ample (effective) Cartier divisor is defined locally on $U$ by a single regular function, this means that the regular functions on $U$ must form a unique factorization domain.
When $\tcA$ is smooth, all ample divisors are Cartier, and we say that $\tcA$ is \textit{factorial}. However, if $\tcA$ contains terminal singularities (i.e. is quasi-smooth), an ample divisor may only be $\mathbb{Q}$-Cartier. In this case, we say that $\tcA$ is $\mathbb{Q}$-\textit{factorial}. 
A variety with only terminal singularities is already a normal variety, so that the regular functions on $U$ are integrally closed, however $\mathbb{Q}$-factoriality is a stronger condition. A hypersurface in a variety of this kind will be smooth~\cite{Cutkosky}

Because reflexive polytopes correspond one-to-one with birational equivalence classes of Gorenstein toric Fano varieties, we are guaranteed that $X=-K_{\cA}$ is already ample, and so we need not introduce any exceptional divisors (discrepancies) to the canonical divisor in the desingularization $\pi$, i.e., $K_{\tcA}=\pi^{*}(K_{\cA})$, and we say that the desingularization is \textit{crepant}~\cite{Batyrev:1994hm,Gross2003}.
As a result, the desingularized space $\tcA$ will still be a Gorenstein toric Fano variety, and therefore projective.

Following Batyrev~\cite{Batyrev:1994hm}, we define a \textit{maximal projective crepant partial (MPCP) desingularization} $\pi :\;\tcA\rightarrow\cA$ to be one such that the pullback $\pi^{*}$ is crepant, and the desingularized space $\tcA$ is $\mathbb{Q}$-factorial and has no worse than terminal singularities.
Furthermore, given any Gorenstein toric Fano variety $\cA$, there exists at least one such MPCP desingularization $\pi$ (see~\cite{Batyrev:1994hm} for the proof).

More importantly for our purposes is how this desingularization is reflected in the polytope formulation.
The removal of non-terminal singularities can be approached by refining the open cover $\mathcal{U}(\tcA)$ on $\tcA$ as much as possible.
Because each maximal 4-cone of $\Delta^{*}$ corresponds to a coordinate patch $U\in\mathcal{U}(\tcA)$, this amounts to subdividing the maximal cones as much as possible such that each subdivision is still a convex rational polyhedral cone with a vertex at the origin.
We call this a \textit{fine, star subdivision} with \textit{star center} at the origin.
The condition of $\mathbb{Q}$-factoriality requires that each new subdivided maximal cone is \textit{simplicial}, i.e., each has four generating rays (since we require $n=4$).

This kind of subdivision is in fact a \textit{triangulation} into simplexes.
Also, because $\tcA$ must be projective, these simplicial cones must be projections of cones from an embedding space.
In the literature, these are referred to as \textit{regular triangulations}~\cite{Gelfand,Lee1991,Thomas2006}.
Thus, in order to find an MPCP desingularization for $\cA$, we must find at least one fine, star, regular triangulation (FSRT) of $\Delta^{*}$.

Before we move on, there is an important point to be made here.
All lattice points other than the origin are either vertices $\mathcal{V}(\Delta^{*})$, or they are not $\hat{\mathcal{V}}(\Delta^{*})=\Delta^{*}\setminus\mathcal{V}(\Delta^{*})$.
Because $\Delta^{*}$ is reflexive and therefore contains no interior points save the origin, it must be true that both $\mathcal{V}(\Delta^{*}),\hat{\mathcal{V}}(\Delta^{*})\subset\partial\Delta^{*}$.
Before desingularization, the generating rays of the maximal cones $\sigma\in\Sigma_{4}(\Delta^{*})$ are the minimal cones $\sigma\in\Sigma_{1}(\Delta^{*})$ whose lattice points include only the origin and points in $\mathcal{V}(\Delta^{*})$.
But, the notion of subdivision of maximal cones implies that before desingularization, $\hat{\mathcal{V}}(\Delta^{*})\neq\emptyset$.
In general, this is true; for each facet $F\in\mathcal{F}_{3}(\Delta^{*})$, there may be lattice points on the boundary such that $\text{skel}_{2}(\Delta^{*})\supset\partial F\neq\emptyset$ and there may be points in the interior such that $\text{skel}_{3}(\Delta^{*})\supset\text{relint}(F)\neq\emptyset$.
However, in the case that $\tcA$ has only terminal singularities, we may ignore points in $\text{relint}(F)$ in the process of subdivison.

This can be explained by considering the orbifold group on $\tcA$, whose construction is given by \cite{Batyrev:2005jc} 

\begin{equation}\label{eq:galois}
\tilde{G}\cong N/\Lambda_{n-1}
\end{equation}
where $\Lambda_{d}$ is the lattice generated by $\text{skel}_{d}(\Delta^{*})$. We now apply an important result of Hasse and Nill~\cite{Haase2005}:

\begin{equation}\label{eq:Hasse}
\text{if }n\geq 3,\;\;\Lambda_{n-2}=\Lambda_{n-1}\, .
\end{equation}
Since we are working with $n=4$, this result implies that $\Lambda_{2}=\Lambda_{3}$, and the orbifold group depends only on $\Lambda_{2}$. Therefore, we may effective ignore points which appear in $\text{skel}_{3}(\Delta^{*})$ and not in $\text{skel}_{2}(\Delta^{*})$.

The underlying reason for this is that with $\Delta^{*}$ reflexive, these points interior to facets correspond precisely to the Demazure roots~\cite{Haase2005,Nill2004a,Batyrev:2005jc,Oda1988} for the orbifold automorphism group $\tilde{G}$.
Thus, to maximize computational efficiency, we need only triangulate the point configuration $\text{skel}_{2}(\Delta^{*})$.

In practice, we want this point configuration to be searchable, so we choose the specific ordering of points given by
\begin{equation}\label{eq:P}
\mathcal{P}(\Delta^{*})=\text{sort}(\mathcal{V}(\Delta^{*}))\cup\text{sort}(\text{skel}_{2}(\Delta^{*})\setminus\mathcal{V}(\Delta^{*}))\, .
\end{equation}
Effectively, then, subdivision of the fan will correspond to expanding the set of vertices of $\Delta^{*}$ from $\mathcal{V}(\Delta^{*})$ to $\mathcal{P}(\Delta^{*})$. We will sometimes refer to the points in $\mathcal{P}(\Delta^{*})$ (or just $\mathcal{P}$) as the \textit{resolved vertices} of $\Delta^{*}$.
Furthermore, vertices $\bm{n}_{\rho}\in\mathcal{P}$ correspond one-to-one with toric divisors $D_{\rho}\subset\tcA$ with bijection $\bm{n}_{\rho}\rightarrow D_{\rho}$.

It is possible to enumerate the FSRTs of the configuration $\mathcal{P}$ (with star center at the origin) using \texttt{TOPCOM}~\cite{Braun2011,Rambau}, however, this becomes highly inefficient as $\text{card}(\mathcal{P})$ becomes large.
A better way to proceed, which is also inherently parallelizable, is to instead consider the configurations $\mathcal{P}\cap\sigma$ for each $\sigma\in\Sigma_{4}(\Delta^{*})$.
The FSRTs of each maximal cone $\sigma$ can then be obtained separately, each using Volker Braun's tremendously useful implementation of \texttt{TOPCOM} in {\tt Sage}, and then recombined.
For each maximal cone, \texttt{TOPCOM} returns a set
\begin{equation}
\mathcal{T}(\mathcal{P}\cap\sigma)=\{T_{\sigma}\;\vert\; T_{\sigma}\text{ an FSRT of } \mathcal{P}\cap\sigma\}\, .
\end{equation}
%

The trade-off for computation efficiency here is that the recombination of triangulated maximal cones is somewhat intricate and tricky.
We use the following algorithm\footnote{We would like to thank Volker Braun for suggesting to us this method of parallelization via the triangulation of maximal cones.}\footnote{A similar algorithm was in use concurrently by Long, McAllister, and McGuirk (see~\cite{Long:2014fba}).}:

{\it
\begin{enumerate}
\item
Choose a triangulation $T_{\sigma}\in\mathcal{T}(\mathcal{P}\cap\sigma)$ for each maximal cone $\sigma\in\Sigma_{4}(\Delta^{*})$.

If all combinations have previously been checked, terminate.
\item
Split up all maximal cones into pairs $(\sigma,\sigma')$ (if there is an odd number, there will be one unpaired cone). 
\item
For one  pair $(\sigma,\sigma')$, check that:
\begin{itemize}
\item
For each simplex $S\in T_{\sigma}$, there exists a simplex $S'\in T_{\sigma'}$ such that $S\cap S'\cap\sigma\cap\sigma'\neq\emptyset$.
\item
For each simplex $S'\in T_{\sigma'}$, there exists a simplex $S\in T_{\sigma}$ such that $S\cap S'\cap\sigma\cap\sigma'\neq\emptyset$.
\item
$T_{\sigma}\cup T_{\sigma'}$ is a regular triangulation (see below).
\end{itemize}
True if $(\sigma,\sigma')$ satisfies all conditions, false otherwise.
\begin{itemize}
\item
If true, repeat step 3 for the next pair.
\item
If false, repeat step 1 with a different combination of triagulations $T_{\sigma}\in\mathcal{T}(\mathcal{P}\cap\sigma)$ for each maximal cone $\sigma\in\Sigma_{4}(\Delta^{*})$.
\end{itemize}
\item
\begin{itemize}
\item
If there is only one pair $(\sigma,\sigma')$, then the triangulation $T=T_{\sigma}\cup T_{\sigma'}$ is an FSRT of $\mathcal{P}$, and therefore of $\Delta^{*}$ (i.e., $T\in\mathcal{T}(\mathcal{P})$).

Repeat step 1 with a different combination of triangulations $T_{\sigma}\in\mathcal{T}(\mathcal{P}\cap\sigma)$ for each maximal cone $\sigma\in\Sigma_{4}(\Delta^{*})$.
\item
Otherwise, define new cones and triangulations by combining each pair via $\tilde{\sigma}=\sigma\cup\sigma'$ and $T_{\tilde\sigma}=T_{\sigma}\cup T_{\sigma'}$.

Split up all new cones into pairs $(\tilde{\sigma},\tilde{\sigma}')$ and repeat step 3.
\end{itemize}
\end{enumerate}
}

To check whether a triangulation $T$ of the point set $\mathcal{P}$ is regular, we use the following well-known algorithm \cite{Lee1991,Billera1990,Thomas2006}

{\it
\begin{enumerate}
\item
Compute the \textit{Gale transform}\footnote{The Gale transform $\mathcal{P}^{\vee}$ of a set of points $\mathcal{P}$ is given by constructing the set of augmented vectors $\hat{\mathcal{P}}=\{(1,\bm{n})\;\vert\;\bm{n}\in\mathcal{P}\}$, and solving the matrix equation $[\hat{\mathcal{P}}]\cdot\left[\mathcal{P}^{\vee}\right]^{T}=\bm{0}$.} $\mathcal{P}^{\vee}$ of $\mathcal{P}$.

\item
For each simplex $S\in T$, define the set $\mathcal{Q}(S)=\{\bm{n}_{i}^{\vee}\in\mathcal{P}^{\vee}\;\vert\; \bm{n}_{i}\in\mathcal{P}\setminus S\}$.

\item
If $\bigcap\limits_{S\in T}{\textrm{{\rm relint}}(\textrm{{\rm cone}}(\mathcal{Q}(S)))}\neq\emptyset$, then $T$ is a regular triangulation of $\mathcal{P}$.

\end{enumerate}
}

\subsection{Weight Matrix}\label{sec:weightmatrix}

Recall from equation (\ref{eq:toric1}) the definition of a toric variety
\begin{equation}
\cA\cong \frac{V}{(C^{*})^{k-n}\times G}\, .
\end{equation}
After desingularization, we obtain a similar toric variety given by
\begin{equation}\label{eq:tildetoric}
\tcA\cong \frac{\tilde{V}}{(C^{*})^{k-n}\times \tilde{G}}\, .
\end{equation}

The group $\tilde{G}$ is nothing more than the orbifold group $\tilde{G}=N/\Lambda_{n-1}$.
However, we must still describe the action on $\tcA$ of the split $(k-n)$-torus $(\mathbb{C}^{*})^{k-n}$ given by the product of 1-tori $\mathbb{C}^{*}$.

The toric variety $\tcA$ may be treated as a weighted projective space with respect to each of the $k-n$ split 1-tori $\mathbb{C}^{*}$ with weights $\bm{w}_{r}=(w_{r}^{\; 1},...,w_{r}^{\; k})\in (\mathbb{Z}_{\geq 0})^{k}$ such that
\begin{equation}
(z_{1},...,z_{k})\sim (\lambda^{w_{r}^{\; 1}}z_{1},...,\lambda^{w_{r}^{\; k}}z_{k}),\;\;\lambda\in\mathbb{C}^{*}\, ,
\end{equation}
for all $r=1,...,k-n$ running over the 1-tori.

It can be shown that the weights $w^{\; \rho}_r$ satisfy an equation of the form
\begin{equation}
\sum_{\rho=1}^{k}{w_{r}^{\;\rho}\cdot \langle\bm{m},\bm{n}_{\rho}\rangle}=0,\;\forall \; \bm{m}\in\Delta\, ,
\end{equation}
or because the fan $\Sigma(\Delta)$ is complete, we can equivalently write
\begin{equation}\label{eq:wn}
\sum_{\rho=1}^{k}{w_{r}^{\;\rho}\,\bm{n}_{\rho}}=\bm{0}\, .
\end{equation}

Recall that for $\tcA$, the vertices of $\Delta^{*}$ are given by $\bm{n}_{\rho}\in\mathcal{P}$ such that $k=\text{card}(\mathcal{P})$.
Then, equation~(\ref{eq:wn}) can be written in matrix form such that
\begin{equation}
[\mathcal{P}]\cdot\bm{W}^{T}=\bm{0}\AND{with}\sum_{\rho=1}^{k}{w_{r}^{\;\rho}}>0\AND{and} \bm{W}\geq 0\, ,
\end{equation}
where $\bm{W}$ is the $(k-n)\times k$ \textit{weight matrix}
\begin{equation}\label{eq:W}
\bm{W}=\left(\begin{array}{c}
\bm{w}_{1}\\
\vdots\\
\bm{w}_{k-n}
\end{array}\right)=\left(\begin{array}{ccc}
w_{1}^{\; 1} & \cdots & w_{1}^{\; k}\\
\vdots & \ddots & \vdots\\
w_{k-n}^{\; 1} & \cdots & w_{k-n}^{\; k}
\end{array}\right)\\ \, .
\end{equation}
Thus, we see that
\begin{equation}
\bm{w}_{1},...,\bm{w}_{k-n}\in\text{ker}([\mathcal{P}])\text{ linearly independent}\, .
\end{equation}
where we assume the kernel to be taken over the integers, so that the weights are well-defined as exponents in a polynomial.

The kernel $\text{ker}([\mathcal{P}])$ can be easily computed in {\tt Sage} given $\mathcal{P}$.
However, in general, non-negativity of the entries of $\bm{W}$ is not guaranteed this way.
We overcome this limitation by restricting to the positive orthant $(\mathbb{Z}_{\geq 0})^{k}$, such that
\begin{equation}
\bm{w}_{1},...,\bm{w}_{k-n}\in\text{ker}([\mathcal{P}])\cap (\mathbb{Z}_{\geq 0})^{k}\text{ linearly independent}\, .
\end{equation}

To perform this computation in {\tt Sage}, we create two \texttt{Polyhedron} objects, one generated by \textit{lines} specified by the elements of $\text{ker}\left([\mathcal{P}]\right)$, and the other generated by \textit{rays} specified by the unit basis vectors:
\begin{lstlisting}
sage: kerPolyhedron = Polyhedron(lines=ker.basis());
sage: posOrthant = Polyhedron(rays=identity_matrix(ker.ngens()).columns());
\end{lstlisting}
The rays of the intersection of these objects are guaranteed to be non-negative, however there may be redundant elements.
Because there are $k-n$ tori, we should find that $\text{rank}(\bm{W})=k-n$.
Again, in the interest of organization, we sort the elements in ascending order and choose the first $k-n$ linearly independent ones.
These will be the rows of the weight matrix $\bm{W}$.

\subsection{The Chow Group and Intersection Numbers}

Next, we review how to compute the Chow group $A^{1}(\tcA)$, which describes the intersection of divisors on $\tcA$.
Recall from Appendix \ref{appa} that the Chow group of Cartier divisors is given by the quotient $A^{1}(\tcA)\cong\mathcal{C}(\tcA)/\sim_{lin}$.
We will therefore need to work out the ideal which generates linear equivalence classes among the divisors.

\subsubsection{Linear Ideal and Stanley--Reisner Ideal}

There is an analogous equation to~(\ref{eq:wn}), the defining equation of the weight matrix, relating toric divisors.
It can be written
\begin{equation}\label{eq:lin}
\sum_{\rho=1}^{k}{\bm{n}_{\rho}\cdot D_{\rho}}=\bm{0}\, .
\end{equation}
This gives a linear relation between the divisors and we define the \textit{linear ideal}
\begin{equation}\label{eq:Ilin}
I_{lin}=\sum_{\rho=1}^{k}{\bm{n}_{\rho}\cdot D_{\rho}}\, .
\end{equation}

In addition, there are non-linear relationships among the toric divisors.
Consider a case where we have $d$ toric divisors such that  $D_{i_{1}}\cdot ...\cdot D_{i_{d}}=\int_{\tcA}{\gamma(D_{i_{1}})\wedge ...\wedge\gamma(D_{i_{d}})}=0$.
In the polytope construction, these divisors with null intersection correspond to the points $\bm{n}_{i_{1}},...,\bm{n}_{i_{d}}\in\mathcal{P}$ which do not appear together as vertices of any simplex $S\in T(\tcA)$ in the FSRT $T(\tcA)$ corresponding to the MPCP desingularization $\tcA$.
The set of these null intersections forms another ideal
\begin{equation}\label{eq:ISR}
I_{SR}(\tcA)=\{D_{i_{1}}\cdot ...\cdot D_{i_{d}}\;\vert\; \bm{n}_{i_{1}},...,\bm{n}_{i_{d}}\not\in  S,\;\forall \;S\in T(\tcA)\}
\end{equation}
known as the \textit{Stanley--Reisner ideal}.
These sets of divisors with null intersections clearly provide another constraint on the Chow group of intersections.

\subsubsection{Chow Group}\label{sec:chowgroup}

Given $I_{lin}$ and $I_{SR}(\tcA)$, we have all the information we need to define linear equivalence classes between the Cartier divisors in $\mathcal{C}(\tcA)$.
Then, we are in a position to define the Chow group as the quotient
\begin{equation}
A^{1}(\tcA)\cong\frac{\mathcal{C}(\tcA)}{I_{lin}+I_{SR}(\tcA)}\, .
\end{equation}

In practice, we do not know what $\mathcal{C}(\tcA)$ is.
However, the Picard group of a toric variety is given by $\text{Pic}(\tcA)\cong\mathbb{Z}^{k-n}$, and therefore we know that $\mathcal{C}(\tcA)/\sim_{lin}\cong\mathbb{Z}^{k-n}$ as well.
If we express $A^{1}(\tcA)$ as a polynomial ring $A_{poly}^{1}(\tcA)$, we can write
\begin{equation}
A_{poly}^{1}(\tcA)\cong\frac{\mathbb{Z}[J_{1},...,J_{k-n}]}{I_{SR}(\tcA)}\, .
\end{equation}

We also do not know what the basis elements $J_{1},...,J_{k-n}$ are in terms of the toric divisor classes $D_{1},...,D_{k}$. However, we can still determine the Chow group using the toric divisor classes and linear equivalence such that
\begin{equation}\label{eq:chowpoly}
A_{poly}^{1}(\tcA)\cong\frac{\mathbb{Z}[D_{1},...,D_{k}]}{I_{lin}+I_{SR}(\tcA)}\, .
\end{equation}

By comparing equations (\ref{eq:wn}) and (\ref{eq:lin}), we see that there is a correspondence\footnote{In fact, the row space of $\bm{W}$ is identical to that of the Mori cone matrix (compare equation (\ref{eq:wn}) to the algorithm used to compute the Mori cone matrix in Section \ref{sec:morikahler2}).} between the columns $\bm{W}^{i}$ of the weight matrix and the toric divisor classes $D_{i}$. We may therefore choose a ``basis'' of divisor classes $\tilde{J}_{1},...,\tilde{J}_{k-n}$ by picking a set of $k-n$ orthogonal columns of $\bm{W}$. However, this ``basis'' is not guaranteed to be orthonormal, and will therefore only span $H^{1,1}(\tcA)$ if given rational coefficients. We then say that $\tilde{J}_{1},...,\tilde{J}_{k-n}$ is a $\mathbb{Q}$-basis.

Like the weight matrix itself, this $\mathbb{Q}$-basis is resolution-independent and therefore is valid for all desingularizations $\tcA$ of $\cA$. This is essential, since it will enable us to accurately compare the Chern classes and intersection numbers of different desingularizations without introducing an arbitrary change of basis. The importance of this property will become clear in Section \ref{sec:wall} when we use Wall's theorem to glue together the various phases of the complete K\"ahler cone corresponding to a distinct Calabi--Yau threefold geometry.

Given the toric divisor classes, a $\mathbb{Q}$-basis of divisor classes, and the linear and Stanley--Reisner ideals, the computation of $A_{poly}^{1}(\tcA)$ in equation (\ref{eq:chowpoly}) is easily accomplished in \texttt{Sage} by defining a \texttt{PolynomialRing} object and an ideal. Because we will be working with a $\mathbb{Q}$-basis of $H^{1,1}(\tcA)$ rather than a $\mathbb{Z}$-basis (see Section \ref{sec:morikahler2}), we must define our polynomial ring $A_{poly}^{1}(\tcA)$ over $\mathbb{Q}$ as follows:
\begin{lstlisting}
sage: C = PolynomialRing(QQ,names=['t']+['D'+str(i+1) for i in range(k)]
+['J'+str(i+1) for i in range(k-n)]);
sage: DD = list(C.gens()[1:-(k-n)]);
sage: JJ = list(C.gens()[-(k-n):]);
sage: ChowIdeal = C.ideal(Ilin+ISR);
\end{lstlisting}

Again, however, in practice we still do not know the $\mathbb{Z}$-basis $J_{1},...,J_{k-n}$.
Later, we will be able to choose one explicitly by considering the K\"ahler cone constraint (see Section \ref{sec:morikahler2}).

\subsubsection{Intersection Numbers}\label{sec:intnum}

Previously, we computed the Chow group $A_{poly}^{1}(\tcA)$ as the quotient group of a polynomial ring.
However, in this construction, the product of elements can only take the form of polynomials.
But as we know, the product of elements of the Chow group is actually an intersection product of 1-cocyles, and not a polynomial product.
Moreover, the intersection product of $n$ 1-cocyles on the $n$-dimensional space $\tcA$, is just an integer in $A^{n}(\tcA)\subset\mathbb{Z}$ (or a rational number in $\mathbb{Q}$ if we have chosen a $\mathbb{Q}$-basis).
Thus, we must choose a normalization for the polynomial ring such that
\begin{equation}
\text{norm}:\; A_{poly}^{n}(\tcA)\overset{\sim}{\rightarrow} A^{n}(\tcA)
\end{equation}
is a bijection.
One such normalization choice involves the lattice volume $\text{vol}(S)$ of a simplex $S\in T(\tcA)$.

If the coordinate patch $U$ has no terminal singularities, i.e., corresponding to points interior to facets on $\Delta^{*}$ (see Section \ref{sec:MPCP}), then the corresponding simplex $S_{U}$ has no such interior points, and we say that it is \textit{elementary}.
Therefore, because all the cones of a reflexive polytope have lattice distance 1, $S_{U}$ must have unit volume, $\text{vol}(S_{U})=1$.
If, however, the coordinate patch $U$ has terminal (i.e., orbifold) singularities, then there will be points interior to the facets of $S_{U}$ (see Section \ref{sec:MPCP}), and the volume of the corresponding simplex will have $\text{vol}(S_{U})>1$.

If $\tcA$ is smooth everywhere, i.e., has no terminal singularities, then the normalization is simple, and every intersection of $n$ 1-cocycles is equal to 1.

Specifically, we define the normalization as follows.
Choose a set of $n$ toric divisor classes $\hat{D}_{1},...,\hat{D}_{n}$ such that they have corresponding vertices $\hat{\bm{n}}_{i_{1}},...,\hat{\bm{n}}_{i_{n}}\in\mathcal{P}\cap \hat{S}$ for $\hat{S}\in T(\tcA)$ a simplex.
Then, for any set of $n$ toric divisor classes $D_{i_{1}},...,D_{i_{n}}$ corresponding to vertices $\bm{n}_{i_{1}},...,\bm{n}_{i_{n}}\in \mathcal{P}\cap S$ for $S\in T(\tcA)$, the normalization takes the form
\begin{equation}\label{eq:norm}
\text{norm}:\;D_{i_{1}}\cdot ...\cdot D_{i_{n}}\mapsto \frac{1}{\text{vol}(\hat{S})}\frac{D_{i_{1}}\cdot ...\cdot D_{i_{n}}}{\hat{D}_{1}\cdot ...\cdot\hat{D}_{n}}\, .
\end{equation}
A Calabi--Yau hypersurface in this construction is defined to be $X=-K_{\tcA}=\sum_{\rho=1}^{k}{D_{\rho}}$ (see Appendix~\ref{appa}).
Since it is a hypersurface, it has codimension 1, and we can therefore find the intersection numbers in the Chow group $A^{n}(X)$ by taking $n-1$ toric divisors $D_{i_{1}},...,D_{i_{n-1}}$, and intersecting them with $X$ directly, i.e., $D_{i_{1}}\cdot ...\cdot D_{i_{n-1}}\cdot X$.
Because $X$ is a formal sum of toric divisor classes, we can still use the same normalization condition in equation (\ref{eq:norm}).

\subsubsection{Favorability}\label{sec:favor}

It is important to note that the toric divisor classes on $\tcA$ do not always descend to a Calabi--Yau hypersurface $X$.
In order to visualize this, consider the short exact sequence
\begin{equation}
0\rightarrow TX\rightarrow T\tcA\vert_{X}\rightarrow\mathcal{N}_{X/\tcA}\rightarrow 0
\end{equation}
with dual sequence
\begin{equation}
0\rightarrow\mathcal{N}^{*}_{X/\tcA}\rightarrow T^{*}\tcA\vert_{X}\rightarrow T^{*}X\rightarrow 0 \, .
\end{equation}
This induces the long exact sequence in sheaf cohomology, part of which is given by
\begin{equation}\label{eq:favor}
\begin{tikzpicture}[descr/.style={fill=white,inner sep=1.5pt}]
        \matrix (m) [
            matrix of math nodes,
            row sep=1em,
            column sep=2.5em,
            text height=1.5ex, text depth=0.25ex
        ]
        {\cdots & H^{1}(X,\mathcal{N}^{*}_{X/\tcA}) & H^{1}(X,T^{*}\tcA\vert_{X}) & H^{1}(X,T^{*}X) \\
            & H^{2}(X,\mathcal{N}^{*}_{X/\tcA}) & H^{2}(X,T^{*}\tcA\vert_{X}) & \cdots \\
        };
        
        \path[overlay,->, font=\scriptsize,>=latex]
        (m-1-1) edge (m-1-2)
        (m-1-2) edge node[descr,yshift=1ex]  {$\alpha$} (m-1-3)
        (m-1-3) edge (m-1-4)
        (m-1-4) edge[out=355,in=175,black] (m-2-2)
        (m-2-2) edge node[descr,yshift=1.2ex] {$\beta$} (m-2-3)
        (m-2-3) edge (m-2-4);
\end{tikzpicture}
\end{equation}
By Dolbeault's theorem, $H^{1}(X,T^{*}X)\cong H^{1,1}(X)\cong A^{1}(X)$.
Then, by the exactness of equation~(\ref{eq:favor}), we find
\begin{equation}
A^{1}(X)\cong\text{coker}(\alpha)\oplus\text{ker}(\beta)\, .
\end{equation}

The cokernel of the map $\alpha$ describes the descent of the K\"ahler moduli on $\tcA$ to K\"ahler moduli on $X$, while the kernel of the map $\beta$ describes ``new" K\"ahler moduli on $X$ which do not descend from $\tcA$.
As long as $\text{ker}(\beta)=0$ and $\text{coker}(\alpha) =H^{1}(X,T^{*}\tcA\vert_{X})$, then all of the K\"ahler forms descend from the ambient space, and we know $A^{1}(X)$ completely.
Otherwise we are missing important information about $A^{1}(X)$.
We then say that $X$, and by a slight abuse of terminology, also the ambient variety $\tcA$ are \textit{unfavorable}.
Studying these unfavorable cases is a problem we leave for future work.
In the present study we simply flag these ambient varieties as unfavorable in the database.

If $X$ is favorable then $\text{dim}(A^{1}(X))\cong\text{dim}(A^{1}(\tcA))$.
This is equivalent to $h^{1,1}(X)=\text{dim}(H^{1,1}(X))\cong\text{dim}(\text{Pic}(\tcA))$.
However, for a toric variety $\text{Pic}(\tcA)=\mathbb{Z}^{k-n}$.
Thus, if $h^{1,1}(X)\neq k-n$, then $\tcA$ is unfavorable.

\subsection{Mori and K\"ahler Cones}\label{sec:morikahler2}

In order to be sure that the hypersurface $X$ is Calabi--Yau, we must ensure that its linear equivalence class $\LD{X}$ is a K\"ahler class, or equivalently that the cohomology class $\gamma(X)$ of its Poincar\'e dual is a K\"ahler form.
This amounts to determining whether $\gamma(X)$ lies within the K\"ahler cone
\begin{equation}\label{eq:kahlercone}
\mathcal{K}(\tcA)=\left\{\left.\omega\in H^{1,1}(\tcA)\;\right\rvert\; \int_{\ND{C}}\!\!\!\!\!\!{\omega}\geq 0,\; \ND{C}\in\overline{\text{NE}}(\tcA)\right\}\, ,
\end{equation}
where $\text{NE}(\tcA)\subset N_{1}(\tcA)$ is the Mori cone (or the cone of (numerically effective) curves)
\begin{equation}\label{eq:moricone}
\text{NE}(\tcA)=\left\{\left.\sum_{i}{a_{i}\ND{C^{i}}}\in N_{1}(\tcA)\;\right\rvert\; a_{i}\in\mathbb{R}_{>0}\right\}=\text{cone}\left(\left\{\ND{C^{i}}\right\}\right)\, ,
\end{equation}
and where $\ND{C^{i}}$ are the numerical equivalence classes of the irreducible, proper curves on $\tcA$.
In practice, we specify these curves via their intersections with the toric divisor classes $D_{1},...,D_{k}\subset\tcA$.
These intersections form a matrix, which we call the \textit{Mori cone matrix}
\begin{equation}\label{eq:moriconemat}
\bm{M}^{i}_{\;\; j}=\ND{C^{i}}\cdot\ND{D_{j}}=\int_{\ND{C^{i}}}\!\!\!\!\!{\gamma(D_{j})}\, .
\end{equation}

Note that the Mori cone itself can be reconstructed from the rows of $\bm{M}$, i.e., \\
$\text{cone}\left(\left\{\bm{M}^{1},...,\bm{M}^{k-n}\right\}\right)\cong\text{NE}(\tcA)$.
Then, the rows of $\bm{M}$ represent the generating curves $\ND{C^{i}}$ of the Mori cone.
In order to calculate these, we use an algorithm originally put forward by Oda and Park~\cite{Oda1988}, though the following version is due to Berglund, Katz, and Klemm~\cite{Berglund:1995gd} (see also~\cite{Grimm:2011fx} and~\cite{Reffert:2007im}):
{\it
\begin{enumerate}
\item
Augment each $\bm{n}_{\rho}\in\mathcal{P}$ to a vector one dimension higher via $\bm{n}_{\rho}\mapsto\bar{\bm{n}}_{\rho}=(1,\bm{n}_{\rho})$.
\item
Find all pairs of $n$-dimensional simplexes $S_{i},S_{j}\in T(\tcA)$ such that $S_{i}\cap S_{j}$ is an $(n-1)$-dimensional simplex, and define the set $\mathcal{S}=\{(S_{i},S_{j})\}$.
\item
For each such pair $s\in\mathcal{S}$, find the unique linear relation $\sum_{\rho=1}^{k}{b_{\rho}^{s}\cdot\bar{\bm{n}}_{\rho}}=0$, such that
\begin{enumerate}[label*=\arabic*.]
\item
All the coefficients $b_{\rho}^{s}$ are minimal integers.
\item
$b_{\rho}^{s}=0$ for $\bm{n}_{\rho}\in\mathcal{P}\setminus (S_{i}\cup S_{j})$, where $s=(S_{i},S_{j})$.
\item
$b_{\rho}^{s}\geq 0$ for $\bm{n}_{\rho}\in (S_{i}\cup S_{j})\setminus (S_{i}\cap S_{j})$, where $s=(S_{i},S_{j})$.
\end{enumerate}
\item
Find a basis of minimal integer vectors $\bm{b}^{s_{1}},...,\bm{b}^{s_{k-n}}$ such that $\bm{b}^{s}$ can be expressed as a positive linear combination, for all $s\in\mathcal{S}$.
\item
The Mori cone matrix is given by $\bm{M}=\left(\begin{array}{c}(\bm{b}^{s_{1}})^{T}\\\vdots\\(\bm{b}^{s_{k-n}})^{T}\end{array}\right)$.
\end{enumerate}
}

We see from equation~(\ref{eq:moriconemat}) that the rows of $\bm{M}$ represent the curves which generate the Mori cone.
Next, we compute the matrix dual to $\bm{M}$ whose columns generate the K\"ahler cone.
We first choose a basis of divisor classes $J_{1},...,J_{k-n}$.
We want these to be the generators of the K\"ahler cone, so from the definition of the K\"ahler cone in equation~(\ref{eq:kahlercone}), they must satisfy
\begin{equation}\label{eq:Kcond}
\int\limits_{\ND{C}}\!\!\!{\gamma(J_{j})}\geq 0,\hspace{5mm}\ND{C}\in\overline{\text{NE}}(\tcA)\, .
\end{equation}
But by the definition of the Mori cone in equation~(\ref{eq:moricone}), we can write
\begin{equation}
\int\limits_{\ND{C}}\!\!\!{\gamma(J_{j})}=\!\!\!\!\!\!\int\limits_{\sum_{i}{a_{i}\ND{C^{i}}}}\!\!\!\!\!\!{\gamma(J_{j})}=\sum_{i}{a_{i}\cdot\left(\int\limits_{\ND{C^{i}}}\!\!\!{\gamma(J_{j})}\right)},\hspace{5mm} a_{i}\in\mathbb{R}_{>0}
\end{equation}
for $\ND{C^{i}}$ the curves generating the Mori cone.

We see then that if equation~(\ref{eq:Kcond}) is satisfied for the generating curves $\ND{C^{i}}$, then it must be satisfied for all $\ND{C}\in\overline{\text{NE}}(\tcA)$.
Thus, we can define the \textit{K\"ahler cone matrix}
\begin{equation}\label{eq:kahlerconemat}
\bm{K}^{i}_{\; j}=\!\!\!\int\limits_{\ND{C^{i}}}\!\!\!{\gamma(J_{j})}\AND{with}\bm{K}\geq 0\, .
\end{equation}
But, because $J_{1},...,J_{k-n}$ form a basis of $A^{1}(\tcA)$, we require the columns of $\bm{K}$ to be orthonormal such that $\bm{K}^{i}_{\; j}=\delta^{i}_{\; j}$.
Comparing equations~(\ref{eq:moriconemat}) and~(\ref{eq:kahlerconemat}), we can then determine the basis $J_{1},...,J_{k-n}$ in terms of the toric divisor classes $D_{1},...,D_{k}$.
Writing the Mori and K\"ahler cone matrices in terms of their columns
\begin{equation}
\bm{M}=(\bm{m}_{1}\;\cdots\;\bm{m}_{k})\AND{and}\bm{K}=(\bm{k}_{1}\;\cdots\;\bm{k}_{k-n})\, ,
\end{equation}
we see that
\begin{equation}\label{eq:kahlerinteger}
J_{j}=\sum_{i=1}^{k}{b_{j}^{\; i}\cdot D_{i}}\AND{when}\bm{K}^{i}_{\; j}=\delta^{i}_{\; j}=\sum_{\rho=1}^{k}{b_{j}^{\;\rho}\bm{M}^{i}_{\;\rho}}\, .
\end{equation}
There are, in general, many choices of such a basis.

The Lefschetz theorem on (1,1)-classes tells us that $A^{1}(\tcA)\cong\textrm{Pic}(\tcA)\cong H^{1,1}(\tcA)\subset H^{2}(\tcA,\mathbb{Z})$.
Therefore, using the above construction ensures that $J_{1},...,J_{k-n}$ generate the K\"ahler cone with integer coefficients (i.e. a $\mathbb{Z}$-basis).
This is an important point if we wish to construct holomorphic line bundles in $\textrm{Pic}(\tcA)$.
Otherwise, however, it is sufficient to construct a basis $\tilde{J}_{1},...,\tilde{J}_{k-n}$ which generates the K\"ahler cone with rational coefficients (a $\mathbb{Q}$-basis).
This relaxes the constraint on the columns of $\bm{K}$ from orthonormality to orthogonality (see Section \ref{sec:chowgroup}).
In this case, we can always choose the basis elements $\tilde{J}_{i}$ to be a subset of the toric divisor classes and define a modified K\"ahler cone matrix $\tilde{\bm{K}}$ such that
\begin{equation}\label{eq:kahlerreal}
\tilde{J}_{j}=\sum_{i=1}^{k}{\delta_{j}^{\; i}\cdot D_{i}}\AND{when}\tilde{\bm{K}}^{i}_{\; j}=\sum_{\rho=1}^{k}{\delta_{j}^{\;\rho}\bm{M}^{i}_{\;\rho}}\, .
\end{equation}
The K\"ahler cone matrix $\tilde{\bm{K}}$ in this case is no longer orthonormal (only orthogonal), and the K\"ahler cone itself no longer trivial. $\mathbb{Z}$- and $\mathbb{Q}$-bases coincide when $\tcA$ is smooth (i.e. factorial).

\subsection{Gluing of K\"ahler Cones}\label{sec:gluing}

We have seen in Section \ref{sec:MPCP} that each Gorenstein toric Fano variety $\cA$ corresponding to a reflexive polytope in the Kreuzer--Skarke database has at least one, but potentially many MPCP desingularizations $\tcA$, and that these correspond exactly to FSRT subdivisions of the fan.
It is not always the case, however, that these desingularizations contain distinct Calabi--Yau hypersurfaces $X$.
Rather, each desingularization $\tcA_{i}$ yields a distinct K\"ahler cone in the K\"ahler moduli space within which the Poincar\'e dual $\gamma(X_{i})$ is constrained.

If the Calabi--Yau hypersurfaces of two or more desingularizations share certain key topological invariants, then it can be shown that they are topologically equivalent and can be considered representations of the same Calabi--Yau threefold.
In this case, the K\"ahler form of this Calabi--Yau threefold is allowed to reside within the K\"ahler cone of either representation, and we refer to these disjoint K\"ahler cone chambers as its \textit{phases}.

In order to allow the K\"ahler form to smoothly vary over its full range, the phases of the K\"ahler cone must be glued together in an appropriate manner.
Because the K\"ahler cone is dual to the Mori cone, this is equivalent to the less intricate task of taking the intersection of the Mori cones corresponding to each K\"ahler cone phase.
This procedure yields a new Mori cone
\begin{equation}\label{eq:moriglue}
\text{NE}(\tcA)=\bigcap_{i}{\text{NE}(\tcA_{i})}\, .
\end{equation}
The new Mori cone matrix is then given by $\bm{M}=\left[\text{rays}\left(\text{NE}(\tcA)\right)\right]^{T}$.
If the Chow group $A^{1}(\tcA_{i})$ of each phase is written in the same basis, then by duality the K\"ahler cone can be determined using either equation (\ref{eq:kahlerinteger}) or (\ref{eq:kahlerreal}) depending on whether it is a $\mathbb{Z}$-basis or a $\mathbb{Q}$-basis (see Sections \ref{sec:chowgroup} and \ref{sec:morikahler2}).

It remains to determine whether some subset of the desingularizations $\tcA_{i}$ of $\cA$ contain hypersurfaces $X_{i}$ which are representations of the same Calabi--Yau threefold $X$.
In the next two sections, we present two, presumably equivalent, methods of determining the full, composite K\"ahler cone corresponding to a distinct Calabi--Yau threefold $X$.

\subsubsection{Chern Classes and Wall's Theorem}\label{sec:wall}

Viewing the Calabi--Yau hypersurfaces $X_{i}$ as real, $2(n-1)$-dimensional (in our case $n=4$) oriented manifolds, we may use an influential theorem due to Wall \cite{Wall1966}:

\vspace{0.5cm}
{\bf Theorem 1.} {\it The homotopy types of complex compact 3-folds are classified by the Hodge numbers, the intersection numbers, and the first Pontryagin class.} \newline

However, because we are working with Calabi--Yau threefolds we can replace the first Pontryagin class with the second Chern class.

The total Chern class of a vector bundle $V$ is given by $c(V)=\sum_{i}{c_{i}(V)}$, where $c_{0}=1$.
In the special case that $V$ is actually a line bundle, then $c_{1}(V)$ is the only non-trivial Chern class, and $c(V)=1+c_{1}(V)$.
The splitting principle tells us that we can break up the Chern class of the vector bundles of interest into the product of Chern classes of line bundles.
Recall that the Gorenstein toric Fano variety $\tcA$ has $k$ toric divisors $D_{1},...,D_{k}$, each of which corresponds to a line bundle $\mathcal{O}_{\tcA}(D_{1}),...\mathcal{O}_{\tcA}(D_{k})$.
As $\tcA$ is 4-dimensional, its Chern class can be written
\begin{equation}\label{eq:cTA1}
c(T\tcA)=1+c_{1}(T\tcA)+c_{2}(T\tcA)+c_{3}(T\tcA)+c_{4}(T\tcA)
\end{equation}

The Chern classes of the ambient variety $\tcA$ can be calculated easily in {\tt Sage} again using the \texttt{PolynomialRing} object implemented earlier in Section~\ref{sec:chowgroup}:
\begin{lstlisting}
sage: cA = prod([(1+C.gen(0)*D) for D in DD]).reduce(ChowIdeal);
sage: cAList = [cA.coefficient({C.gen(0):i})
for i in range(cA.degree(C.gen(0)))];
\end{lstlisting}

In order to calculate the Chern classes of the Calabi--Yau hypersurface $X$, we consider the short exact sequence
\begin{equation}
0\rightarrow TX\rightarrow T\tcA\vert_{X}\rightarrow \mathcal{N}_{X/\tcA}\vert_{X}\cong\mathcal{O}_{\tcA}(X)\vert_{X}=\mathcal{O}_{X}(X)\rightarrow 0\, .
\end{equation}
Recall that $c_{1}(\mathcal{O}_{X}(X))=X$.
By the definition of the Chern class and because $X$ is 3-dimensional, we write
\begin{align}\label{eq:cTA2}
c(T\tcA\vert_{X})&=c(TX)\; c(\mathcal{O}_{X}(X))=\Big(1+c_{1}(TX)+c_{2}(TX)+c_{3}(TX)\Big)\Big(1+X\Big) \\
&=1+\Big(c_{1}(TX)+X\Big)+\Big(c_{2}(TX)+c_{1}(TX)\,X\Big)+\Big(c_{3}(TX)+c_{2}(TX)\,X\Big)+\ldots\; .\notag
\end{align}
%

However, the Calabi--Yau condition tells us that $c_{1}(TX)=0$ and thus, comparing equations (\ref{eq:cTA1}) and (\ref{eq:cTA2}), we find that $c_{1}(T\tcA)=X$ and therefore, after some algebra, that
\begin{align}
c_{2}(TX)=c_2(T\tcA)\AND{and} c_{3}(TX)=c_{3}(T\tcA)-c_{2}(T\tcA)\; c_{1}(T\tcA) \;.
\end{align}

Furthermore, the Euler number calculated in Section \ref{sec:hodge} can be checked by integrating the top Chern class
\begin{equation}
\chi(X)=\int\limits_{X}{c_{3}(TX)}
\end{equation}

We now have enough information to compute all of the information required by Wall's theorem.
If these quantities are identical for multiple desingularizations $\tcA_{i}$ of $\cA$, then their hypersurfaces $X_{i}$ should be considered identical and their K\"ahler cone phases glued via equations~(\ref{eq:moriglue}), (\ref{eq:kahlerinteger}), and (\ref{eq:kahlerreal}).

\subsubsection{Identifying Flop Transitions}

A presumably\footnote{We say ``presumably" here due to the following issue. Wall's theorem is enough to ensure that two 3-folds are equivalent as real manifolds. However, it may be that the natural complex structure inherited from the ambient space for two different descriptions of a Calabi--Yau threefold are not in the same connected component of complex structure moduli space.} equivalent method of determining when the K\"ahler cone phases of two desingularizations $\tcA_{i}$ should be glued amounts to checking whether or not all singularities in the walls between these phases are avoided by the Calabi--Yau hypersurface. This can be done by tracing a curve with negative self-intersection through a \textit{flop} in the wall of the K\"ahler cone and making sure that the Calabi--Yau hypersurface misses this curve on both sides.

A curve $C$ with self-intersection $C^{2}<0$ necessarily has negative intersection with every toric divisor which contains it, i.e., $C\cdot D_{j_{1}}<0$, ..., $C\cdot D_{j_{d}}<0$ for $C\subset D_{j_{1}}\cap ...\cap D_{j_{d}}$.
If the transition between two phases is a flop, then such a curve $C$ will blow down to a point on the K\"ahler cone wall, and then blow up again to a new curve $C'$ on the adjacent phase.
Then, we can use the following algorithm due to Berglund, Katz, Klemm, and Mayr~\cite{Berglund:1996uy} (see also~\cite{Grimm:2011fx}):

{\it
\begin{enumerate}
\item
Compare the Mori cone matrices $\bm{M}_{1}$ and $\bm{M}_{2}$ of two phases corresponding to two desingularizations $\tcA_{1}$ and $\tcA_{2}$ of $\cA$.

If a row of $\bm{M}_{1}$ appears in $\bm{M}_{2}$ with its signs flipped, then these rows represent generators $C_{1}$ and $C_{2}$ of the Mori cone $\text{NE}(\tcA_{1})$ and $\text{NE}(\tcA_{2})$, such that $C_{1}$ which blows down to a point in the wall of the K\"ahler cone and then blows up to $C_{2}$  on the other side.

Equivalently, there exists a flop from $\tcA_{1}$ to $\tcA_{2}$.

\item
Determine a subvariety $V_{1}=D_{i_{1}}\cap\cdots\cap D_{i_{d}}$ ($d<n$) of $\tcA_{1}$ from the intersection of toric divisors which have negative intersection with $C_{1}$ (i.e., columns $i_{1},...,i_{d}$ of $\bm{M}_{1}$ with negative entries on the row corresponding to $C_{1}$).

Similarly, determine a subvariety $V_{2}=D_{i_{1}}\cap\cdots\cap D_{i_{d}}$ ($d<n$) of $\tcA_{2}$ from the intersection of toric divisors which have negative intersection with $C_{2}$ (i.e., columns $i_{1},...,i_{d}$ of $\bm{M}_{2}$ with negative entries on the row corresponding to $C_{2}$).

\item
If $V_{1}\cdot X_{1}=0$ and $V_{2}\cdot X_{2}=0$, then $X=X_{1}=X_{2}$ is a single Calabi--Yau hypersurface, and the flop does not exist in $X$.

\item
Repeat steps 1 through 3 for all pairs of adjacent K\"ahler cone phases corresponding to desingularizations of $\cA$.

\item
Each group of desingularizations $\{\tcA_{i}\}$ which are related by flops that do not exist in the hypersurface defines a single Calabi--Yau geometry $X$.

The Mori cone of $X$ is obtained by taking the intersection of the Mori cones of the associated desingularizations via equations~(\ref{eq:moriglue}).
\end{enumerate}
}
In practice, however, we have used the gluing procedure based on Wall's theorem discussed in Section \ref{sec:wall}.

\section{Querying the Database: Illustrative Examples}\label{sec:database}

Max Kreuzer and Harald Skarke have compiled a complete database of all reflexive polytopes in four dimensions.
Our aim here is to provide a catalogue of geometrical properties of as many of the associated Calabi--Yau threefold geometries as possible.

The Kreuzer--Skarke database catalogs Newton polytopes which encode the ambient toric varieties in which Calabi--Yau threefolds are embedded as hypersurfaces.
Each Newton polytope has a dual, the triangulations of which correspond to separate Calabi--Yau geometries.
In this manner, there exists a vastly larger quantity of Calabi--Yau threefolds than reflexive polytopes.
Frequently, however, some subset of the triangulations of a dual polytope encode identical topological information, the major difference being the content of the K\"ahler cone.
In such cases, we must treat these as chambers or \textit{phases}, which we ``glue together'' into a larger K\"ahler cone corresponding to a single Calabi--Yau geometry.

In the interest of full clarity, the following example database entry will focus on a polytope whose triangulations are divided into two distinct geometries, one of which has a K\"ahler cone with multiple phases.
For those familiar with the Kreuzer--Skarke database, this will be the $95^{\text{th}}$ reflexive polytope in dimension $4$ with $h^{1,1}=3$.

\subsection{Search Fields}\label{sec:SEARCHBASIC}

There are currently two ways to access the database of toric Calabi--Yau threefolds. The first is simple, but limited to a series of text fields and checkboxes, and the second allows the user to enter complex custom querying commands in \texttt{SQL}.

\subsubsection{Basic Query}

\begin{figure}[!h]
\centering
\includegraphics[width=0.98\textwidth]{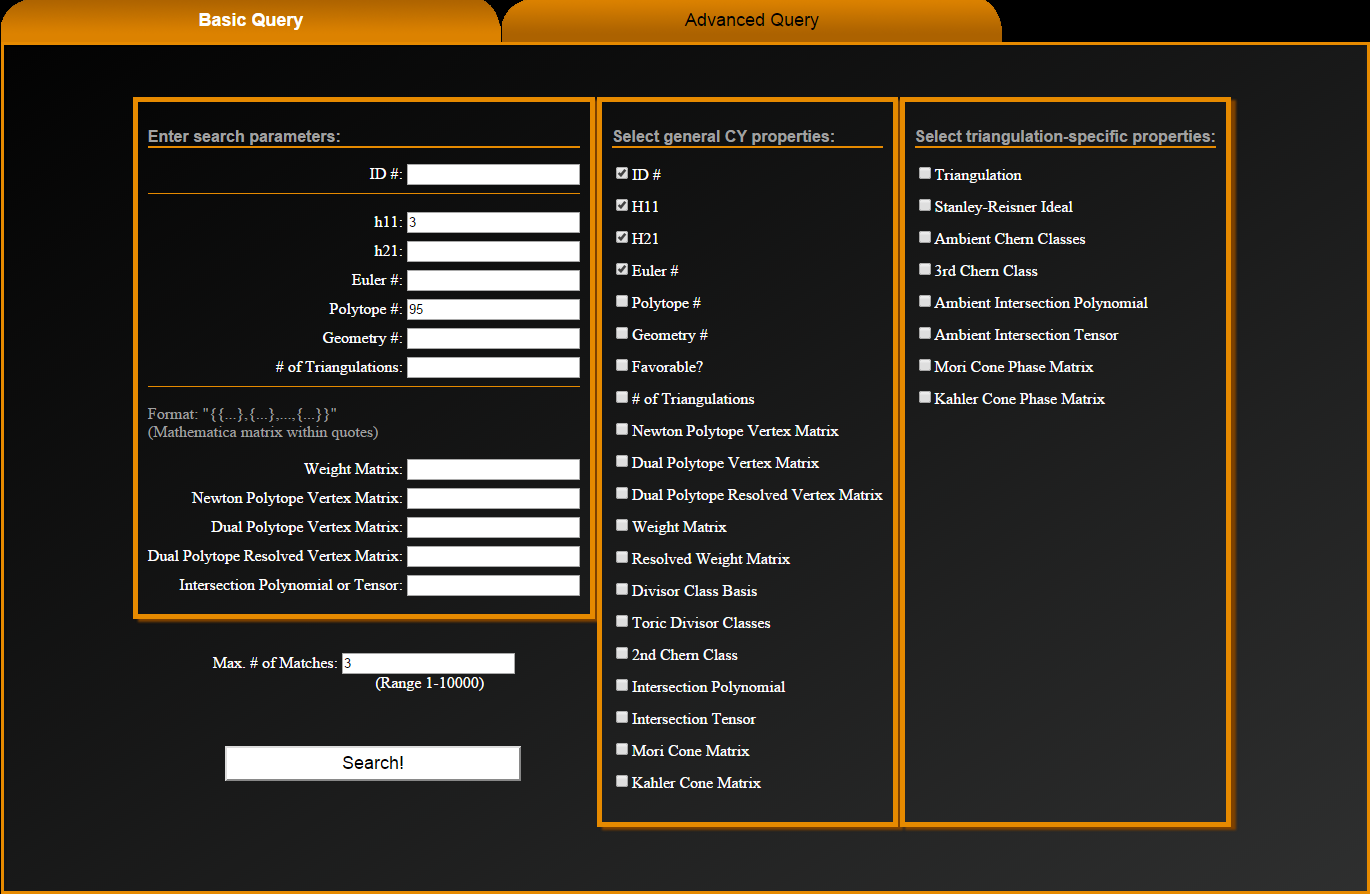}
\caption{Field-Based Search}
\label{fig:SEARCHBASIC}
\end{figure}

There are many different uses for a database such as this, and therefore we provide a variety of ways to sift through it.
The left hand box in Figure \ref{fig:SEARCHBASIC} allows the user to enter the parameters of the search.
The user may choose to specify all, some, or none of the search parameters.
In the case that none are specified, entries will be returned in the order in which they were first entered, starting at the beginning of the database.
The user may specify a maximum number of matches between 1 and 10,000 for search results.
This field is set to 1 by default.

Also, the user my search for multiple values of any field by entering them sequentially in the input box separated by commas.
The allowed search fields are:

\begin{itemize}
\item
\textbf{ID \#}: If this field is specified, at most a single entry will be returned since each Calabi--Yau threefold is indexed with a unique ID.
For this reason, \textbf{ID \#} should not be specified unless the user knows exactly which geometry to isolate.
\end{itemize}

The next block of fields are all integer-valued:
\begin{itemize}
\item
\textbf{h11}: The Hodge number $h^{1,1}=\text{dim }H^{1,1}(X)$.

\item
\textbf{h21}: The Hodge number $h^{2,1}=\text{dim }H^{2,1}(X)$.

\item
\textbf{Euler \#}: The Euler number $\chi (X)=2(h^{1,1}-h^{2,1})$.

\item
\textbf{Polytope \#}: The index of a polytope in the Kreuzer--Skarke database for a given value of $h^{1,1}$.

\item
\textbf{Geometry \#}: Since each polytope may give rise to multiple Calabi--Yau geometries, we index by these as well, for example in the following way:\\\\

{\centering
\begin{tikzpicture}[auto,
every node/.style={text width=5em, text centered, minimum height=1.5cm },
node distance=6cm]

\tikzstyle{level 1}=[level distance=3cm, sibling distance=2.5cm]
\tikzstyle{pg}=[rectangle, draw,
edge from parent path={(\tikzparentnode.south) -- (\tikzchildnode.north)}]

\begin{scope}[grow=down, sloped, edge from parent/.style={draw=black!70,-latex}]
\node[pg] {Polytope 1}
    child { node[pg] {Geometry 1} }
    child { node[pg] {Geometry 2} }
;
\end{scope}
\begin{scope}[grow=down, sloped, edge from parent/.style={draw=black!70,-latex}, xshift=3.75cm]
\node[pg] {Polytope 2}
    child { node[pg] {Geometry 3} }
;
\end{scope}
\begin{scope}[grow=down, sloped, edge from parent/.style={draw=black!70,-latex}, xshift=8.75cm]
\node[pg] {Polytope 3}
    child { node[pg] {Geometry 4} }
    child { node[pg] {Geometry 5} }
    child { node[pg] {Geometry 6} }
;
\end{scope}
\begin{scope}[xshift=13.25cm, edge from parent/.style={draw=none}]
\node {$\cdots$}
    child { node {$\cdots$} }
;
\end{scope}
\end{tikzpicture}
\newline
\par}

\item
\textbf{\# of Triangulations}: Recall that triangulations of a polytope which result in identical topological structure must be glued together to form the K\"ahler cone for that geometry.
This parameter is the number of such triangulations that glue together to form the geometry in question.
\end{itemize}

The final block of fields are all String-valued and must be enclosed in quotation marks.
Each of these must be formatted as a Mathematica matrix (or tensor) with no spaces:

\begin{itemize}
\item
\textbf{Weight Matrix}: This is the weight matrix (see Section \ref{sec:weightmatrix}) of the ambient toric variety (viewed as a weighted projective space with multiple sets of weights) in which the Calabi--Yau threefold in question is embedded.
Kreuzer and Skarke refer to this in the literature as a CWS or \textit{combined weight system}.\\

This field will search for matches to weight matrices before and after desingularization of the ambient toric variety.\\

\textit{Note: The rows of the weight matrix must be sorted in ascending order before searching the database.}\\

See Sections \ref{sec:CWS} and \ref{sec:RESCWS} for an example of proper formatting.

\item
\textbf{Newton Polytope Vertex Matrix}: This is the matrix of a Newton polytope appearing in the Kreuzer--Skarke database.
When searching, its rows and columns must be in the same order as they appear there.\\

See Section \ref{sec:NVERTS} for an example of proper formatting.

\item
\textbf{Dual Polytope Vertex Matrix}: This is the matrix of the dual to a Newton polytope appearing in the Kreuzer--Skarke database.\\

Because the ordering of its vertices can be ambiguous, please be sure to sort them (i.e., the columns of the matrix) in ascending order before searching.\\

See Section \ref{sec:DVERTS} for an example of proper formatting.

\item
\textbf{Dual Polytope Resolved Vertex Matrix}: This is the vertex matrix $\mathcal{P}$ (see Section \ref{sec:MPCP}) of the dual polytope after subdivision.
Some of the vertices given by the columns of this matrix are not necessary to define the convex hull of the polytope, and ignoring them leaves us with the Dual Polytope Vertex Matrix of the previous entry.
We call the full set of vertices, including these ``extra vertices'', \textit{resolved vertices}.\\

Again, because the ordering of the non-interior points is ambiguous, please use the Dual Polytope Vertex Matrix augmented on the right by the ``extra vertices'', which should be sorted in ascending order (see equation (\ref{eq:P})).

See Section \ref{sec:DRESVERTS} for an example of proper formatting.

\item
\textbf{Intersection Polynomial or Tensor}: This is the triple-intersection number tensor, which is a topological invariant of the Calabi--Yau threefold in question.
It should be written in Mathematica notation as a nested array of size $h^{1,1}\times h^{1,1}\times h^{1,1}$.
Because the intersection tensor is fully symmetric, the ordering is \textit{not} ambiguous in this case, although of course the basis choice can be.\\

It is also possible to use as input the intersection numbers in polynomial form with divisor class basis elements as variables.
However, this introduces unnecessary ambiguities in ordering, and we recommend that the user work with tensors.\\

See Section \ref{sec:IPOLY} and \ref{sec:ITENS} for examples of proper formatting in each case.
\end{itemize}

\subsubsection{Advanced Query}

\begin{figure}[!h]
\centering
\includegraphics[width=0.98\textwidth]{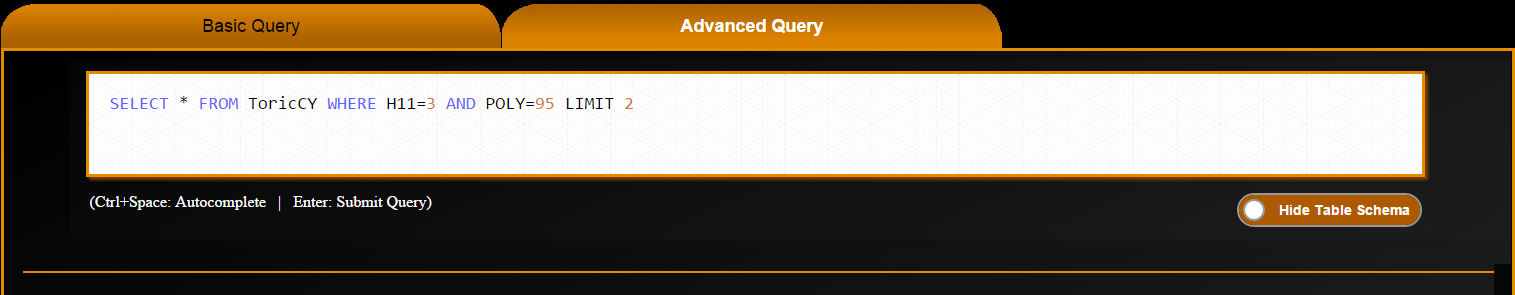}
\caption{SQL-Based Search}
\label{fig:SEARCHADVANCED}
\end{figure}

This option provides the user familiar with \texttt{SQL} with a command prompt, allowing him/her to enter a customized query. The above query returns all properties of each total Calabi--Yau geometry corresponding to polytope 95 in the Kreuzer--Skarke list with $h^{1,1}=3$.

By untoggling the button labelled ``Hide Table Schema'', the user may view the available properties and their column names in the \texttt{SQL} table as shown below

\begin{figure}[!h]
\centering
\includegraphics[width=0.98\textwidth]{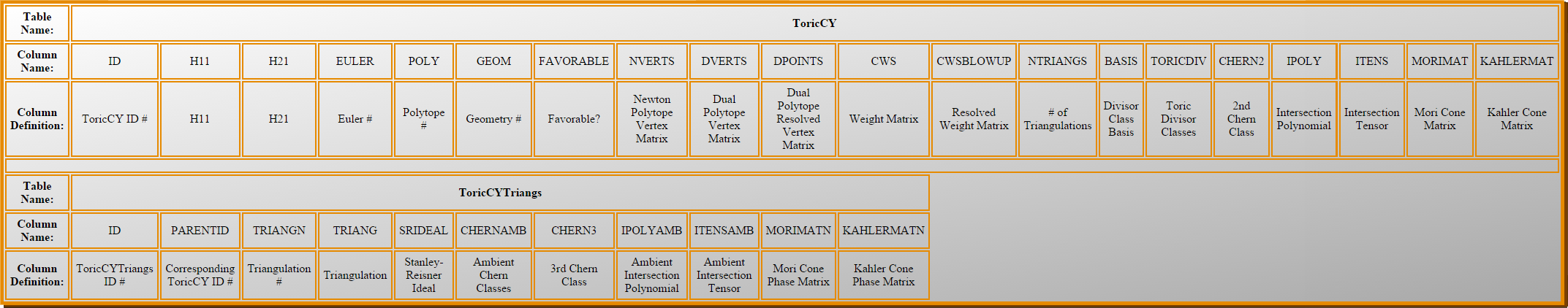}
\caption{Properties and Column Names in \texttt{SQL} Tables}
\label{fig:TABLESCHEMA}
\end{figure}

\subsection{Search Results for General Calabi--Yau Properties}\label{sec:gencyprop}

\subsubsection{The Essentials: \textbf{ID \#}, \textbf{H11}, \textbf{H21}, and \textbf{Euler \#}}\label{sec:ID-EULER}

\begin{figure}[H]
\centering
\includegraphics[height=90px]{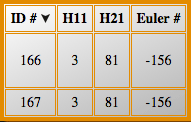}
\label{fig:ID-EULER}
\end{figure}

These four entries will always appear in the results of a database search.
\textbf{ID \#} is the global index within the database of the Calabi--Yau threefold in question.\\

\textbf{H11} and \textbf{H21} are its Hodge numbers, and \textbf{Euler \#} is $\chi (X)=2(h^{1,1}-h^{2,1})$.\\

\textit{* These properties correspond to a particular ambient toric variety, and each embedded geometry will share it.}\\

The two geometries of polytope $95$ with $h^{1,1}=3$ have database ID numbers 166 and 167.
Both geometries have Hodge numbers $h^{1,1}=3$ and $h^{2,1}=81$, and Euler number $-156$.

\subsubsection{\textbf{Polytope \#}}\label{sec:POLY}

\begin{figure}[H]
\centering
\includegraphics[height=90px]{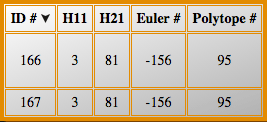}
\label{fig:POLY}
\end{figure}

The \textbf{Polytope \#} is the index of a polytope in the Kreuzer--Skarke database for a given value of $h^{1,1}$.\\

As mentioned, this example uses polytope $95$.\\

\textit{* This property corresponds to a particular ambient toric variety, and each embedded geometry will share it.}

\subsubsection{\textbf{Geometry \#}}\label{sec:GEOM}

\begin{figure}[H]
\centering
\includegraphics[height=90px]{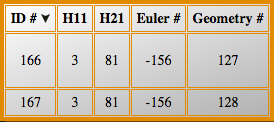}
\label{fig:GEOM}
\end{figure}

The \textbf{Geometry \#} is the index used to specify a particular Calabi--Yau threefold within the set of geometries of a given $h^{1,1}$.
Clusters of these geometries belong to a single polytope in the following way:\\\\

{\centering
\begin{tikzpicture}[auto,
every node/.style={text width=5em, text centered, minimum height=1.5cm },
node distance=6cm]

\tikzstyle{level 1}=[level distance=3cm, sibling distance=2.5cm]
\tikzstyle{pg}=[rectangle, draw,
edge from parent path={(\tikzparentnode.south) -- (\tikzchildnode.north)}]

\begin{scope}[grow=down, sloped, edge from parent/.style={draw=black!70,-latex}]
\node[pg] {Polytope 1}
    child { node[pg] {Geometry 1} }
    child { node[pg] {Geometry 2} }
;
\end{scope}
\begin{scope}[grow=down, sloped, edge from parent/.style={draw=black!70,-latex}, xshift=3.75cm]
\node[pg] {Polytope 2}
    child { node[pg] {Geometry 3} }
;
\end{scope}
\begin{scope}[grow=down, sloped, edge from parent/.style={draw=black!70,-latex}, xshift=8.75cm]
\node[pg] {Polytope 3}
    child { node[pg] {Geometry 4} }
    child { node[pg] {Geometry 5} }
    child { node[pg] {Geometry 6} }
;
\end{scope}
\begin{scope}[xshift=13.25cm, edge from parent/.style={draw=none}]
\node {$\cdots$}
    child { node {$\cdots$} }
;
\end{scope}
\end{tikzpicture}
\newline
\par}

In this example, the geometries resulting from polytope $95$ in $h^{1,1}=3$ have ID numbers 166 and 167 within the full database, but geometry numbers 127 and 128 when restricted to the set with $h^{1,1}=3$.\\

\textit{* This property corresponds to a particular Calabi--Yau threefold geometry.}

\subsubsection{Is the Geometry \textbf{Favorable?}}\label{sec:FAVORABLE}

\begin{figure}[H]
\centering
\includegraphics[height=90px]{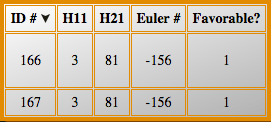}
\label{fig:FAVORABLE}
\end{figure}

As discussed in Section~\ref{sec:favor}, we call a Calabi--Yau hypersurfaces \textit{unfavorable} if the ambient space K\"ahler forms do not descend to provide a basis of $H^{1,1}(X)$.
This search parameter is a Boolean flag which reads 1 for favorable and 0 for unfavorable.\\

In our example, both geometries 166 and 167 are favorable.\\

\textit{* This property corresponds to a particular Calabi--Yau threefold geometry.}

\subsubsection{\textbf{\# of Triangulations}}\label{sec:NTRIANGS}

\begin{figure}[H]
\centering
\includegraphics[height=90px]{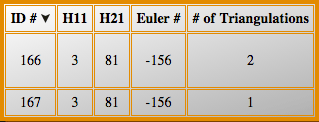}
\label{fig:NTRIANGS}
\end{figure}

The value of this parameter tells us how many topologically-identical triangulations of the parent polytope had their K\"ahler cones glued together to form the moduli space of this Calabi--Yau threefold.\\

In this example, polytope $95$ for $h^{1,1}=3$ has three triangulations, two of which were topologically-equivalent and suitable for gluing.
This resulted in the two distinct Calabi--Yau geometries 166 and 167, one of which is composed of two triangulations, and the other of one triangulation.\\

\textit{* This property corresponds to a particular Calabi--Yau threefold geometry.}

\subsubsection{\textbf{Newton Polytope Vertex Matrix}}\label{sec:NVERTS}

\begin{figure}[H]
\centering
\includegraphics[height=90px]{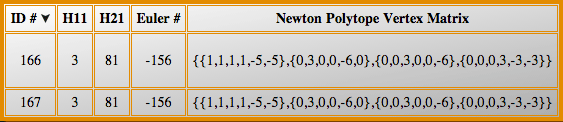}
\label{fig:NVERTS}
\end{figure}

This is the Mathematica-style matrix of a Newton polytope appearing in the Kreuzer--Skarke database.
Its rows and columns are given in the same order in which they appear there.\\

In our example, geometries 166 and 167 both descend from the Newton polytope $\Delta$ on the lattice $M\cong\mathbb{Z}^{4}$ with vertices given by the columns of the matrix

\begin{equation*}
[\mathcal{V}(\Delta)]=\left(
\begin{array}{cccccc}
 1 & 1 & 1 & 1 & -5 & -5 \\
 0 & 3 & 0 & 0 & -6 & 0 \\
 0 & 0 & 3 & 0 & 0 & -6 \\
 0 & 0 & 0 & 3 & -3 & -3
\end{array}
\right)
\end{equation*}

Note that both Newton polytopes are identical.
This is because $\Delta$ defines a particular singular toric variety which is resolved to provide the ambient spaces within which both geometries are hypersurfaces.\\

\textit{* This property corresponds to a particular ambient toric variety, and each embedded geometry will share it.}

\subsubsection{\textbf{Dual Polytope Vertex Matrix}}\label{sec:DVERTS}
\begin{figure}[H]
\centering
\includegraphics[height=90px]{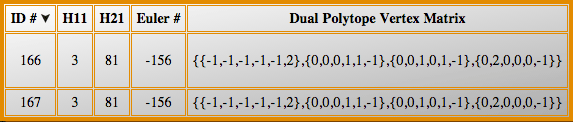}
\label{fig:DVERTS}
\end{figure}

This is the Mathematica-style matrix of the dual to a Newton polytope $\Delta$ appearing in the Kreuzer--Skarke database.\\

Again, in our example, the dual polytope $\Delta^{*}$ corresponding to geometries 166 and 167 has vertices on the lattice $N\cong\mathbb{Z}^{4}$ given by the columns of the matrix

\begin{equation*}
[\mathcal{V}(\Delta^{*})]=\left(
\begin{array}{cccccc}
 -1 & -1 & -1 & -1 & -1 & 2 \\
 0 & 0 & 0 & 1 & 1 & -1 \\
 0 & 0 & 1 & 0 & 1 & -1 \\
 0 & 2 & 0 & 0 & 0 & -1
\end{array}
\right)
\end{equation*}

Note that the vertices (i.e., the columns of the matrix) are sorted in ascending order.\\

\textit{* This property corresponds to a particular ambient toric variety, and each embedded geometry will share it.}

\subsubsection{\textbf{Dual Polytope Resolved Vertex Matrix}}\label{sec:DRESVERTS}
\begin{figure}[H]
\centering
\includegraphics[width=0.98\textwidth]{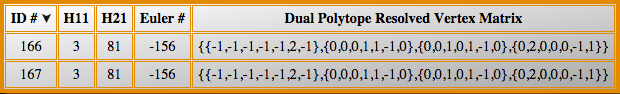}
\label{fig:DRESVERTS}
\end{figure}

The resolved points $\mathcal{P}$ (see Section \ref{sec:MPCP}) of the dual polytope $\Delta^{*}$ are those points which are not in the interior region of any cone in the fan of $\Delta^{*}$.
The vertices of $\Delta^{*}$ are examples of non-interior points, but there may be others.
These can be thought of as ``extra vertices'', which must be taken into account for the purposes of triangulation, but which are redundant in defining the convex hull of $\Delta^{*}$.

The \textbf{Dual Polytope Resolved Vertex Matrix} is just the \textbf{Dual Polytope Vertex Matrix} augmented on the right by the ``extra vertices'', with the later sorted in ascending order (see equation (\ref{eq:P})).\\

In our example, the resolved vertices of the dual polytope $\Delta^{*}$ corresponding to geometries 166 and 167 are given by the columns of the matrix

\begin{equation*}
[\mathcal{P}]=\left(
\begin{array}{ccccccc}
 -1 & -1 & -1 & -1 & -1 & 2 & -1 \\
 0 & 0 & 0 & 1 & 1 & -1 & 0 \\
 0 & 0 & 1 & 0 & 1 & -1 & 0 \\
 0 & 2 & 0 & 0 & 0 & -1 & 1
\end{array}
\right)
\end{equation*}

\textit{* This property corresponds to a particular ambient toric variety, and each embedded geometry will share it.}

\subsubsection{\textbf{Weight Matrix}}\label{sec:CWS}

\begin{figure}[H]
\centering
\includegraphics[height=90px]{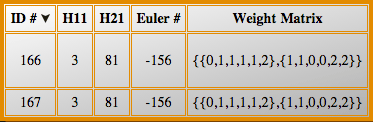}
\label{fig:CWS}
\end{figure}

For each torus action on the ambient variety $\cA$ in which a Calabi--Yau threefold is embedded, there is a set of non-negative weights (see Section \ref{sec:weightmatrix}) in each of the coordinates which define $\cA$ in a manner analogous to weighted projective space.
The \textbf{Weight Matrix} $\bm{W}$ is the matrix for which each row contains the weights for a particular torus action (see equation (\ref{eq:W})).
Kreuzer and Skarke refer to this in the literature as a CWS or \textit{combined weight system}.\\

In this example, the ambient toric variety corresponding to polytope $95$ with $h^{1,1}=3$ has weight matrix

\begin{equation*}
\bm{W}=\left(
\begin{array}{cccccc}
 0 & 1 & 1 & 1 & 1 & 2 \\
 1 & 1 & 0 & 0 & 2 & 2
\end{array}
\right)
\end{equation*}

Note that there are as many weights in each row as there are vertices in the Dual Polytope Vertex Matrix.\\

Also, note that the rows of this matrix are sorted in ascending order.\\

\textit{* This property corresponds to a particular ambient toric variety, and each embedded geometry will share it.}

\subsubsection{\textbf{Resolved Weight Matrix}}\label{sec:RESCWS}

\begin{figure}[H]
\centering
\includegraphics[height=90px]{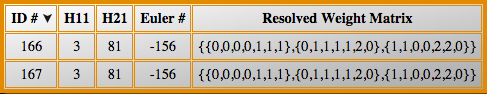}
\label{fig:RESCWS}
\end{figure}

The \textbf{Weight Matrix} $\bm{W}$ describes the ambient toric variety $\cA$ corresponding to a polytope in the Kreuzer--Skarke database.
However, $\cA$ may be singular, and therefore its singularities may need to be resolved to some degree before it can give rise to a smooth Calabi--Yau threefold as a hypersurface.
This resolution is performed by blowing up the singular points of $\cA$, resulting in a new ambient variety $\tcA$ with a new weight matrix $\tilde{\bm{W}}$.\\

In this example, the resolved ambient toric variety corresponds to the weight matrix

\begin{equation*}
\tilde{\bm{W}}=\left(
\begin{array}{ccccccc}
 0 & 0 & 0 & 0 & 1 & 1 & 1 \\
 0 & 1 & 1 & 1 & 1 & 2 & 0 \\
 1 & 1 & 0 & 0 & 2 & 2 & 0
\end{array}
\right)
\end{equation*}

Note that there are as many weights in each row as there are non-interior points in the \textbf{Dual Polytope Resolved Vertex Matrix}.\\

Also, note that the rows of this matrix are sorted in ascending order.\\

\textit{* This property corresponds to a particular ambient toric variety, and each embedded geometry will share it.}

\subsubsection{\textbf{Divisor Class Basis}}\label{sec:BASIS}

\begin{figure}[H]
\centering
\includegraphics[height=90px]{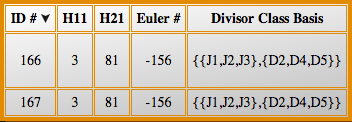}
\label{fig:BASIS}
\end{figure}

Because the Chow group $A^{1}(X)\cong H^{1,1}(X)$, it must have dimension equal to $h^{1,1}$.
Furthermore, since $H^{1,1}(X)$ can be thought of as the vector space of K\"ahler moduli of the Calabi--Yau threefold $X$, any divisor class can be written in terms of $h^{1,1}$ independent basis elements (see Section \ref{sec:morikahler2}).
Our notation for these basis elements with \textit{real coefficients} is specified here.\\

In this example, we already know that $h^{1,1}=3$, so we have the basis elements $J_{1}$, $J_{2}$, and $J_{3}$, which in this case correspond to the three toric divisors $D_{2}$, $D_{4}$, and $D_{5}$ (see equation (\ref{eq:kahlerreal})).\\

\textit{* This property corresponds to a particular ambient toric variety, and each embedded geometry will share it.}

\subsubsection{\textbf{Toric Divisor Classes}}\label{sec:TORICDIV}

\begin{figure}[H]
\centering
\includegraphics[height=90px]{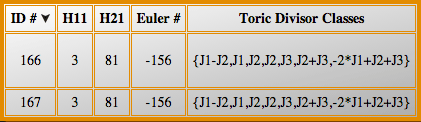}
\label{fig:TORICDIV}
\end{figure}

Each of the non-interior points of a dual polytope $\Delta^{*}$ (or edges of its subdivided fan) corresponds to a divisor class in the desingularized ambient toric variety $\tcA$.
We refer to these as the \textbf{Toric Divisor Classes}, and we express them in terms of the divisor class basis with real coefficients.
Here, we denote each divisor class $\LD{D}$ by a representative $D$.

In this example, the toric divisor classes are given by
\begin{equation*}
\begin{array}{lllll}
D_{1}=J_{1}-J_{2} & & D_{2}=J_{1} & & D_{3}=J_{2}\\
 & & \\
D_{4}=J_{2} & & D_{5}=J_{3} & & D_{6}=J_{2}+J_{3}\\
 & & \\
 & & D_{7}=-2 J_{1}+J_{2}+J_{3} & &
\end{array}
\end{equation*}

Note that there are as many toric divisor classes as there are columns of the \textbf{Dual Polytope Resolved Vertex Matrix}, and one corresponds to the other in the same order.\\

\textit{* This property corresponds to a particular ambient toric variety, and each embedded geometry will share it.}

\subsubsection{\textbf{$2^{\text{nd}}$ Chern Class}}\label{sec:CHERN2}

\begin{figure}[H]
\centering
\includegraphics[height=90px]{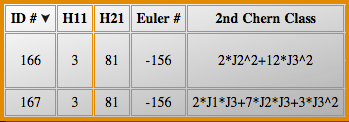}
\label{fig:CHERN2}
\end{figure}

This is the $2^{\text{nd}}$ Chern class $c_{2}(TX)$ specific to the Calabi--Yau threefold $X$ given in terms of the divisor class basis with real coefficients.\\

In this example, $c_{2}(TX_{166})=2 J_{2}^{2}+12 J_{3}^{2}$ and $c_{2}(TX_{167})=2 J_{1}J_{3}+7 J_{2}J_{3}+3 J_{3}^{2}$.\\

\textit{* This property corresponds to a particular Calabi--Yau threefold geometry.}

\subsubsection{\textbf{Intersection Polynomial}}\label{sec:IPOLY}

\begin{figure}[H]
\centering
\includegraphics[height=90px]{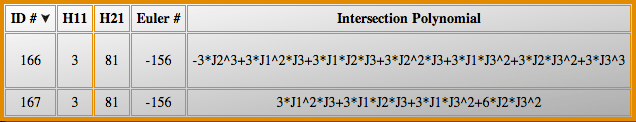}
\label{fig:IPOLY}
\end{figure}

These are the triple-intersection numbers $\kappa_{ijk}(X)$ specific to the Calabi--Yau threefold $X$ written in the compact polynomial notation

\begin{equation*}
P(X)=\sum_{i=1}^{h^{1,1}}{\sum_{j=i}^{h^{1,1}}{\sum_{k=j}^{h^{1,1}}{\kappa_{ijk}(X)\, J_{i}J_{j}J_{k}}}}
\end{equation*}

In this example, we have
\begin{align*}
P(X_{166})&=-3 J_{2}^{3}+3 J_{1}^{2} J_{3}+3 J_{1} J_{2} J_{3}+3 J_{2}^{2} J_{3}+3 J_{1} J_{3}^{2}+3 J_{2} J_{3}^{2}+3 J_{3}^{3}\\
P(X_{167})&=3 J_{1}^{2} J_{3}+3 J_{1} J_{2} J_{3}+3 J_{1} J_{3}^{2}+6 J_{2} J_{3}^{2}
\end{align*}

\textit{* This property corresponds to a particular Calabi--Yau threefold geometry.}

\subsubsection{\textbf{Intersection Tensor}}\label{sec:ITENS}

\begin{figure}[H]
\centering
\includegraphics[height=90px]{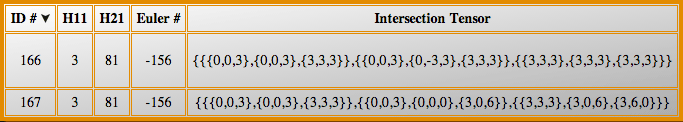}
\label{fig:ITENS}
\end{figure}

These are the triple intersection numbers $\kappa_{ijk}(X)$ specific to the Calabi--Yau threefold $X$ written in tensor form.\\

In this example, we have

\begin{equation*}
\kappa_{1jk}(X_{166})=\left(
\begin{array}{ccc}
 0 & 0 & 3 \\
 0 & 0 & 3 \\
 3 & 3 & 3
\end{array}
\right),\hspace{5mm}
\kappa_{2jk}(X_{166})=\left(
\begin{array}{ccc}
 0 & 0 & 3 \\
 0 & -3 & 3 \\
 3 & 3 & 3
\end{array}
\right),\hspace{5mm}
\kappa_{3jk}(X_{166})=\left(
\begin{array}{ccc}
 3 & 3 & 3 \\
 3 & 3 & 3 \\
 3 & 3 & 3
\end{array}
\right)
\end{equation*}

and

\begin{equation*}
\kappa_{1jk}(X_{167})=\left(
\begin{array}{ccc}
 0 & 0 & 3 \\
 0 & 0 & 3 \\
 3 & 3 & 3
\end{array}
\right),\hspace{5mm}
\kappa_{2jk}(X_{167})=\left(
\begin{array}{ccc}
 0 & 0 & 3 \\
 0 & 0 & 0 \\
 3 & 0 & 6
\end{array}
\right),\hspace{5mm}
\kappa_{3jk}(X_{167})=\left(
\begin{array}{ccc}
 3 & 3 & 3 \\
 3 & 0 & 6 \\
 3 & 6 & 0
\end{array}
\right)
\end{equation*}

\textit{* This property corresponds to a particular Calabi--Yau threefold geometry.}

\subsubsection{\textbf{Mori Cone Matrix}}\label{sec:MORIMAT}

\begin{figure}[H]
\centering
\includegraphics[height=90px]{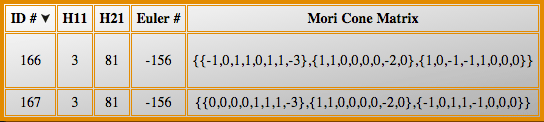}
\label{fig:MORIMAT}
\end{figure}

This is the Mori cone matrix, the rows of which represent a basis of irreducible, proper curves on $\tcA$ (see Section \ref{sec:morikahler2}).
This matrix $\bm{M}$ corresponds to that in equation (\ref{eq:moriconemat}).

The Mori cone of the full Calabi--Yau threefold $X$ is the intersection of the Mori cones of its various phases $\tcA^{p}$, each corresponding to a distinct triangulation.
Each row of $\bm{M}^{i}_{\; j}(\tcA^{p})$ represents a basis element of the set $N_{1}(\tcA^{p})$ of 1-cycles modulo numerical equivalence.
We can build up the Mori cone or ``cone of curves'' from $\bm{M}^{i}_{\; j}(\tcA^{p})$, by treating each row as a vertex coordinate in a vector space.
Taking the intersection of these cones for all phases $p$, we can deconstruct it again by treating its extremal rays as the rows of a matrix.
Keeping only linearly independent rows, this is now the full Mori cone matrix $\bm{M}^{i}_{\; j}(X)$ (see Section \ref{sec:gluing}).\\

In our example, geometries 166 and 167 have the following Mori cone matrices

\begin{equation*}
\bm{M}(\tcA_{166})=\left(
\begin{array}{cccccccc}
 -1 & 0 & 1 & 1 & 0 & 1 & 1 & -3 \\
 1 & 1 & 0 & 0 & 0 & 0 & -2 & 0 \\
 1 & 0 & -1 & -1 & 1 & 0 & 0 & 0
\end{array}
\right)
\end{equation*}
and
\begin{equation*}
\bm{M}(\tcA_{167})=
\left(
\begin{array}{cccccccc}
 0 & 0 & 0 & 0 & 1 & 1 & 1 & -3 \\
 1 & 1 & 0 & 0 & 0 & 0 & -2 & 0 \\
 -1 & 0 & 1 & 1 & -1 & 0 & 0 & 0
\end{array}
\right)
\end{equation*}

Note that there is one more column of the Mori cone matrix than columns of the \textbf{Dual Polytope Resolved Vertex Matrix}.
This is because the final column, in fact, corresponds to the origin and can be ignored for most practical purposes, however it is recorded here for completeness.\\

\textit{* This property corresponds to a particular Calabi--Yau threefold geometry.}

\subsubsection{\textbf{K\"ahler Cone Matrix}}\label{sec:KAHLERMAT}

\begin{figure}[H]
\centering
\includegraphics[height=90px]{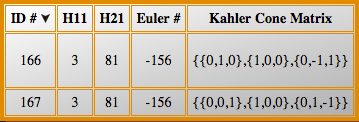}
\label{fig:KAHLERMAT}
\end{figure}

The requirement that a Calabi--Yau threefold $X$ must be an ample divisor in an ambient toric variety $\cA$ puts a strict requirement on the K\"ahler form $\omega$ considered as the Poincar\'e dual of the divisor class of $X$.
In fact, we find that $\omega$ must lie in the K\"ahler cone (see Section \ref{sec:morikahler2}).

When the geometry is favorable, the $h^{1,1}$ divisor class basis elements $J_{i}$ of the desingularized ambient toric variety $\tcA_{i}$ descend to the Calabi--Yau hypersurface $X$, so that the K\"ahler cone is simplicial, and can therefore be expressed via an $h^{1,1}\times h^{1,1}$ matrix of unique intersection numbers $\bm{K}(\tcA_{i})$ (see equation (\ref{eq:kahlerconemat})).
However, the K\"ahler cone will not, in general, be simplicial after gluing multiple phases together and the resultant K\"ahler cone matrix $\bm{K}(\tcA)$ will have at least as many (but potentially more) rows as columns.

When we choose a basis of divisor classes with integer coefficients, the K\"ahler cone of each phase will be trivial with K\"ahler cone matrix $\bm{K}(\tcA_{i})$ equal to an identity matrix.
However, when we choose a basis with real coefficients, this will not always be the case.

Because the \textbf{Divisor Class Basis} in this database has real coefficients, the K\"ahler cone matrix will generally be non-trivial.
Furthermore, these matrices are the end result of gluing phases, and will therefore in general have more rows than columns.

In our example, geometries 166 and 167 have the following K\"ahler cone matrices

\begin{equation*}
\mathcal{K}(\tcA_{166})=\left(
\begin{array}{ccc}
 0 & 1 & 0 \\
 1 & 0 & 0 \\
 0 & -1 & 1
\end{array}
\right)
\end{equation*}
and
\begin{equation*}
\mathcal{K}(\tcA_{167})=
\left(
\begin{array}{ccc}
 0 & 0 & 1 \\
 1 & 0 & 0 \\
 0 & 1 & -1
\end{array}
\right)
\end{equation*}

\textit{* This property corresponds to a particular Calabi--Yau threefold geometry.}

\subsection{Search Results for Triangulation-Specific Calabi--Yau Properties}\label{sec:speccyprop}

This final set of search results are specific to a particular fine, regular, star triangulation (FSRT) of the dual polytope $\Delta^{*}$, and are therefore not of much use for most physical calculations.
However, they may be instructive in other ways, so we include them.

For each FSRT of the dual polytope $\Delta^{*}$, there is a unique resolution $\tcA$ of the singularities of the ambient toric variety $\cA$ encoded therein.
Each desingularized ambient variety contains an embedded Calabi--Yau hypersurface $X$.
Following Wall's Theorem, the Hodge numbers $h^{1,1}(X)$ and $h^{2,1}(X)$, the Euler number $\chi (X)$, the intersection tensor $\kappa_{ijk}(X)$, and the second Chern class $c_{2}(TX)$ are taken to be topological invariants of $X$.
Therefore, any two desingularizations $\tcA^{1}$ and $\tcA^{2}$ whose Calabi--Yau hypersurfaces $X^{1}$ and $X^{2}$ have $h^{1,1}(X^{1})=h^{1,1}(X^{2})$, $h^{2,1}(X^{1})=h^{2,1}(X^{2})$, $\chi (X^{1})=\chi (X^{2})$, $\kappa_{ijk}(X^{1})=\kappa_{ijk}(X^{2})$, and $c_{2}(X^{1})=c_{2}(X^{2})$ are considered to be identical and each of their moduli spaces are taken to be ``phases'' of the total Calabi--Yau threefold.

In this example search result, both geometries 166 and 167 are constructed from a single dual polytope $\Delta^{*}$, which has 3 triangulations.
Each of these triangulations corresponds to a unique desingularization ($\tilde{\mathcal{A}}_{166}^{1}$, $\tcA_{166}^{2}$, or $\tcA_{167}$) of the ambient toric variety $\cA$.
Furthermore, each desingularized variety has embedded a Calabi--Yau hypersurface ($X_{166}^{1}$, $X_{166}^{2}$, and $X_{167}$).
By Wall's criteria, $X_{166}^{1}$ and $X_{166}^{2}$ are topologically equivalent and therefore may be treated as phases of a whole Calabi--Yau geometry $X_{166}$ (geometry 166 in the database).
The remaining hypersurface $X_{167}$ therefore constitutes its own Calabi--Yau geometry (geometry 167 in the database).

\subsubsection{\textbf{Triangulation \#}}\label{sec:TRIANGN}

\begin{figure}[!h]
\centering
\includegraphics[height=90px]{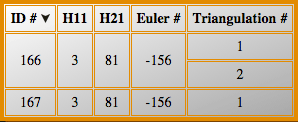}
\label{fig:TRIANGN}
\end{figure}

Within a given geometry, the \textbf{Triangulation \#} indexes the particular triangulation (and therefore the phase of the moduli space) in question.

In this example, geometry 166 is composed of two triangulations indexed by 1 and 2, and geometry 167 is composed of only one triangulation which has the index 1.

\subsubsection{\textbf{Triangulation}}\label{sec:TRIANG}

\begin{figure}[!h]
\includegraphics[width=0.98\textwidth]{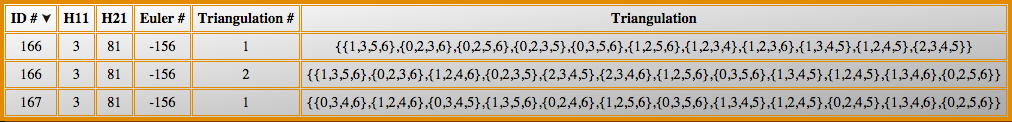}
\label{fig:TRIANG}
\end{figure}

This result tells us precisely how the convex hull of the dual polytope $\Delta^{*}$ is triangulated into simplexes.
Each subarray represents a simplex and its contents are the column indices of the \textbf{Dual Polytope Resolved Vertex Matrix} which correspond to vertices in $\mathcal{P}$ of the simplex.\\

These are triangulations of the convex hull boundary of $\Delta^{*}$, but by placing an extra vertex at the origin $(0,0,0,0)$ in every simplex, they become star triangulations of the full volume of $\Delta^{*}$.\\

In our example, both geometries 166 and 167 have \textbf{Dual Polytope Resolved Vertex Matrix}

\begin{equation*}
[\mathcal{P}]=\begin{blockarray}{ccccccc}
\matindex{\textbf{0}} & \matindex{\textbf{1}} & \matindex{\textbf{2}} & \matindex{\textbf{3}} & \matindex{\textbf{4}} & \matindex{\textbf{5}} & \matindex{\textbf{6}} \\
\begin{block}{(ccccccc)}
 -1 & -1 & -1 & -1 & -1 & 2 & -1 \\
 0 & 0 & 0 & 1 & 1 & -1 & 0 \\
 0 & 0 & 1 & 0 & 1 & -1 & 0 \\
 0 & 2 & 0 & 0 & 0 & -1 & 1 \\
\end{block}
\end{blockarray}
\end{equation*}

Then for geometry 166, the first simplex of both its triangulations contains the vertices

\begin{equation*}
\{1,3,5,6\}\Rightarrow\left\{\left(\begin{array}{c} -1\\ 0\\ 0\\ 2\end{array}\right),\left(\begin{array}{c} -1\\ 1\\ 0\\ 0\end{array}\right),\left(\begin{array}{c} 2\\ -1\\ -1\\ -1\end{array}\right),\left(\begin{array}{c} -1\\ 0\\ 0\\ 1\end{array}\right)\right\}
\end{equation*}

And for geometry 167, the first simplex of its single triangulation contains the vertices

\begin{equation*}
\{0,3,4,6\}\Rightarrow\left\{\left(\begin{array}{c} -1\\ 0\\ 0\\ 0\end{array}\right),\left(\begin{array}{c} -1\\ 1\\ 0\\ 0\end{array}\right),\left(\begin{array}{c} -1\\ 1\\ 1\\ 0\end{array}\right),\left(\begin{array}{c} -1\\ 0\\ 0\\ 1\end{array}\right)\right\}
\end{equation*}

\subsubsection{\textbf{Stanley--Reisner Ideal}}\label{sec:SRIDEAL}

\begin{figure}[!h]
\centering
\includegraphics[height=90px]{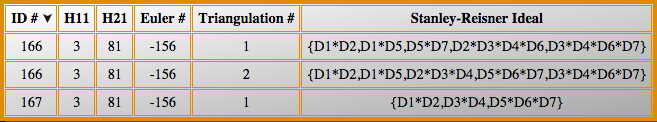}
\label{fig:SRIDEAL}
\end{figure}

This is the Stanley--Reisner ideal $I_{\text{SR}}(\tcA)$, a set of maximal ideals corresponding to a specific desingularization of the ambient toric variety $\tcA$, or equivalently to a specific triangulation of the dual polytope $\Delta^{*}$.
In short, it dictates which subsets of toric divisor classes never intersect at a point.\\

A toric variety is in general a non-trivial algebraic variety, but it can be understood more concretely as a quotient space of $\mathbb{C}^{k}$ (minus some exceptional set and where $k$ is given by the number of toric divisor classes) in a construction very similar to that of a simple projective space.\\

Let $\mathbb{C}^{k}$ have coordinates $(z_{1},...,z_{k})$.
One of the basic properties of a toric variety $\tcA$ is that each of its toric divisor classes $D_{i}$ is given by the subvariety $D_{i}=\{z_{i}=0\}$.
If two toric divisor classes never intersect at a point, then the intersection
\begin{equation*}
D_{1}\cdot D_{2}=\int\limits_{\tcA}{D_{1}\wedge D_{2}\wedge D\wedge E}=0
\end{equation*}
for any arbitrary divisor classes $D$ and $E$ in $\tcA$, and therefore we find $D_{1}\cdot D_{2}\subset I_{\text{SR}}(\tcA)$.
If no other proper subset of $I_{\text{SR}}(\tcA)$ contains $D_{1}\cdot D_{2}$, then it is maximal.
In terms of coordinates, $D_{1}\cdot D_{2}$ defines the subvariety $Z(D_{1}\cdot D_{2})$ by coordinates $(0,0,z_{3},...,z_{k})\in\mathbb{C}^{k}$.\\

Following this procedure, we construct a subvariety $Z({I_{i}})$ for every maximal ideal $I_{i}\subset I_{\text{SR}}(\tcA)$.
Then, removing the union
\begin{equation*}
Z(I_{\text{SR}}(\tcA))=\bigcup\limits_{i}{Z(I_{\text{SR}})}
\end{equation*}
from $\mathbb{C}^{k}$ ensures that the correct sets of toric divisor classes never intersect at a point.

Then, the desingularized ambient toric variety $\tcA$ can be written
\begin{equation*}
\tcA\cong\frac{ \mathbb{C}^{k}\setminus Z(I_{\text{SR}}(\tcA)) }{ (\mathbb{C}^{*})^{k-n}\times\tilde{G}}
\end{equation*}
where the toric group action $(\mathbb{C}^{*})^{k-n}$ is defined by the equivalence relations
\begin{equation*}
(z_{1},...,z_{k})\sim \left(\lambda^{\tilde{\bm{W}}_{i}^{\; 1}}z_{1},...,\lambda^{\tilde{\bm{W}}_{i}^{\; k}}z_{k}\right),\hspace{3mm}\lambda\in \mathbb{C}^{*}, \forall i=1,...,k-n
\end{equation*}
with $\tilde{W}$ the \textbf{Resolved Weight Matrix}.\\

In this example, the two triangulations comprising geometry 166 give rise to the Stanley--Reisner ideals
\begin{align*}
I_{\text{SR}}(\tcA_{166}^{1})&=\{D_{1}\cdot D_{2},D_{1}\cdot D_{5},D_{5}\cdot D_{7},D_{2}\cdot D_{3}\cdot D_{4}\cdot D_{6},D_{3}\cdot D_{4}\cdot D_{6}\cdot D_{7}\}\\
I_{\text{SR}}(\tcA_{166}^{2})&=\{D_{1}\cdot D_{2},D_{1}\cdot D_{5},D_{2}\cdot D_{3}\cdot D_{4},D_{5}\cdot D_{6}\cdot D_{7},D_{3}\cdot D_{4}\cdot D_{6}\cdot D_{7}\}\, .
\end{align*}
The single triangulation comprising geometry 167 gives rise to the Stanley--Reisner ideal
\begin{equation*}
I_{\text{SR}}(\tcA_{167})=\{D_{1}\cdot D_{2},D_{3}\cdot D_{4},D_{5}\cdot D_{6}\cdot D_{7}\}\, .
\end{equation*}

\subsubsection{\textbf{Ambient Chern Classes}}\label{sec:CHERNAMB}

\begin{figure}[!h]
\centering
\includegraphics[width=0.98\textwidth]{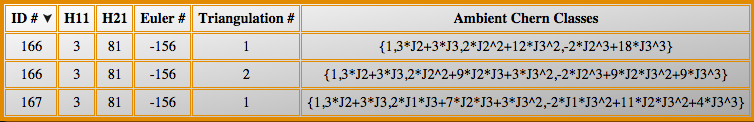}
\label{fig:CHERNAMB}
\end{figure}

These are the Chern classes ($c_{0}(T\tcA)$, $c_{1}(T\tcA)$, $c_{2}(T\tcA)$, and $c_{3}(T\tcA)$) of a desingularized ambient toric variety $\tcA$ given in terms of the divisor class basis.
These Chern classes are not invariants associated to the polytope, and are therefore triangulation-specific.

The two triangulations comprising geometry 166 give rise to ambient Chern classes
\begin{equation*}
\begin{array}{lcl}
c_{0}(T\tcA_{166}^{1})=1 & & c_{1}(T\tcA_{166}^{1})=3 J_{2}+3 J_{3}\\
c_{2}(T\tcA_{166}^{1})=2 J_{2}^{2}+12 J_{3}^{2} & & c_{3}(T\tcA_{166}^{1})=-2 J_{2}^{3}+18 J_{3}^{3}\\
 & & \\
 & \text{and} & \\
 & & \\
c_{0}(T\tcA_{166}^{2})=1 & & c_{1}(T\tcA_{166}^{2})=3 J_{2}+3 J_{3}\\
c_{2}(T\tcA_{166}^{2})=2 J_{2}^{2}+9 J_{2} J_{3}+3 J_{3}^{2} & & c_{3}(T\tcA_{166}^{2})=-2 J_{2}^{3}+9 J_{2} J_{3}^{2}+9 J_{3}^{3}
\end{array}
\end{equation*}

The single triangulation comprising geometry 167 gives rise to ambient Chern classes
\begin{equation*}
\begin{array}{ll}
c_{0}(T\tcA_{167})=1 & c_{1}(T\tcA_{167})=3 J_{2}+3 J_{3}\\
c_{2}(T\tcA_{167})=2 J_{1} J_{3}+7 J_{2} J_{3}+3 J_{3}^{2} & c_{3}(T\tcA_{167})=-2 J_{1} J_{3}^{2}+11 J_{2} J_{3}^{2}+4 J_{3}^{3}
\end{array}
\end{equation*}

\subsubsection{\textbf{$3^{\text{rd}}$ Chern Class}}\label{sec:CHERN3}

\begin{figure}[!h]
\centering
\includegraphics[height=90px]{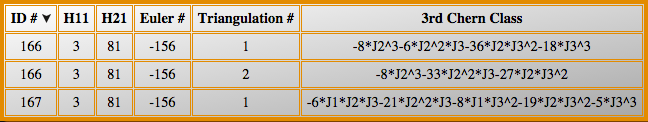}
\label{fig:CHERN3}
\end{figure}

This is the $3^{\text{rd}}$ Chern class $c_{3}(TX)$ of the Calabi--Yau threefold $X$.
Though it might naively seem from these results like it takes different values on the different phases of $X$, they can be shown to be identical when taking into account the triple-intersection numbers.

In this example, the two triangulations comprising geometry 166 give rise to
\begin{align*}
c_{3}(TX_{166}^{1})&=-8 J_{2}^{3}-6 J_{2}^{2} J_{3}-36 J_{2} J_{3}^{2}-18 J_{3}^{3}\\
c_{3}T(X_{166}^{2})&=-8 J_{2}^{3}-33 J_{2}^{2} J_{3}-27 J_{2} J_{3}^{2}
\end{align*}
The single triangulation comprising geometry 167 gives rise to
\begin{equation*}
c_{3}(TX_{167})=-6 J_{1} J_{2} J_{3}-21 J_{2}^{2} J_{3}-8 J_{1} J_{3}^{2}-19 J_{2} J_{3}^{2}-5 J_{3}^{3}
\end{equation*}

In all cases, the different products of the $J_i$ are all proportional to the top form on the threefold.
Using the intersection numbers to compute the constants of proportionality in each case we find that $c_3(TX)$ is indeed an invariant as it should be.
Integrating this over the Calabi--Yau threefold, we obtain the Euler number, $-156$ in this case
\begin{equation*}
\chi(X)=\int\limits_{X}{c_{3}(TX)}
\end{equation*}

\subsubsection{\textbf{Ambient Intersection Polynomial}}\label{sec:IPOLYAMB}

\begin{figure}[!h]
\centering
\includegraphics[width=\textwidth]{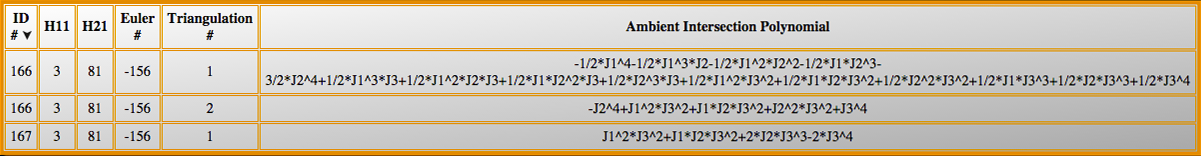}
\label{fig:IPOLYAMB}
\end{figure}

These are the quadruple intersection numbers $\kappa_{ijkl}(\tcA)$ specific to the desingularized ambient toric variety $\tcA$ written in the compact polynomial notation

\begin{equation*}
P(\tcA)=\sum_{i=1}^{h^{1,1}}{\sum_{j=i}^{h^{1,1}}{\sum_{k=j}^{h^{1,1}}{\sum_{l=k}^{h^{1,1}}{\kappa_{ijkl}(X)\, J_{i}J_{j}J_{k}J_{l}}}}}
\end{equation*}

In this example, the two triangulations comprising geometry 166 give rise to

\begin{align*}
P(\tcA_{166}^{1})&=-\frac{1}{2} J_{1}^{4}-\frac{1}{2} J_{1}^{3} J_{2}-\frac{1}{2} J_{1}^{2} J_{2}^{2}-\frac{1}{2} J_{1} J_{2}^{3}-\frac{3}{2} J_{2}^{4}+\frac{1}{2} J_{1}^{3} J_{3}+\frac{1}{2} J_{1}^{2} J_{2} J_{3}+\frac{1}{2} J_{1} J_{2}^{2} J_{3}\\
&\hspace{5mm}+\frac{1}{2} J_{2}^{3} J_{3}+\frac{1}{2} J_{1}^{2} J_{3}^{2}+\frac{1}{2} J_{1} J_{2} J_{3}^{2}+\frac{1}{2} J_{2}^{2} J_{3}^{2}+\frac{1}{2} J_{1} J_{3}^{3}+\frac{1}{2} J_{2} J_{3}^{3}+\frac{1}{2} J_{3}^{4}\\\\
P(\tcA_{166}^{2})&=-J_{2}^{4}+J_{1}^{2} J_{3}^{2}+J_{1} J_{2} J_{3}^{2}+J_{2}^2 J_{3}^{2}+J_{3}^{4}
\end{align*}

The single triangulation comprising geometry 167 gives rise to

\begin{equation*}
P(\tcA_{167})=J_{1}^{2} J_{3}^{2}+J_{1} J_{2} J_{3}^{2}+2 J_{2} J_{3}^{3}-2 J_{3}^{4}
\end{equation*}

\subsubsection{\textbf{Ambient Intersection Tensor}}\label{sec:ITENSAMB}

\begin{figure}[!h]
\centering
\includegraphics[height=90px]{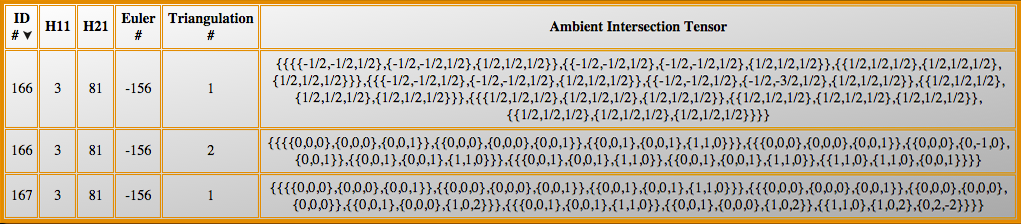}
\label{fig:ITENSAMB}
\end{figure}

These are the quadruple intersection numbers $\kappa_{ijkl}(\tcA)$ specific to the desingularized ambient toric variety $\tcA$ written in tensor form.\\

In this example, the two triangulations comprising geometry 166 give rise to

\begin{equation*}
\kappa_{11kl}(\tcA_{166}^{1})=\left(
\begin{array}{ccc}
 -\frac{1}{2} & -\frac{1}{2} & \frac{1}{2} \\
 -\frac{1}{2} & -\frac{1}{2} & \frac{1}{2} \\
 \frac{1}{2} & \frac{1}{2} & \frac{1}{2}
\end{array}
\right),\hspace{2mm}
\kappa_{22kl}(\tcA_{166}^{1})=\left(
\begin{array}{ccc}
 -\frac{1}{2} & -\frac{1}{2} & \frac{1}{2} \\
 -\frac{1}{2} & -\frac{3}{2} & \frac{1}{2} \\
 \frac{1}{2} & \frac{1}{2} & \frac{1}{2}
\end{array}
\right),\hspace{2mm}
\kappa_{33kl}(\tcA_{166}^{1})=\left(
\begin{array}{ccc}
 \frac{1}{2} & \frac{1}{2} & \frac{1}{2} \\
 \frac{1}{2} & \frac{1}{2} & \frac{1}{2} \\
 \frac{1}{2} & \frac{1}{2} & \frac{1}{2}
\end{array}
\right)
\end{equation*}

\begin{equation*}
\begin{array}{llllll}
\kappa_{12kl}(\tcA_{166}^{1}) & =\kappa_{21kl}(\tcA_{166}^{1}) & \kappa_{13kl}(\tcA_{166}^{1}) & =\kappa_{31kl}(\tcA_{166}^{1}) & \kappa_{32kl}(\tcA_{166}^{1}) & =\kappa_{23kl}(\tcA_{166}^{1})
\\\\
 & =\left(
\begin{array}{ccc}
 -\frac{1}{2} & -\frac{1}{2} & \frac{1}{2} \\
 -\frac{1}{2} & -\frac{1}{2} & \frac{1}{2} \\
 \frac{1}{2} & \frac{1}{2} & \frac{1}{2}
\end{array}
\right), &
& =\left(
\begin{array}{ccc}
 \frac{1}{2} & \frac{1}{2} & \frac{1}{2} \\
 \frac{1}{2} & \frac{1}{2} & \frac{1}{2} \\
 \frac{1}{2} & \frac{1}{2} & \frac{1}{2}
\end{array}
\right), &
& =\left(
\begin{array}{ccc}
 \frac{1}{2} & \frac{1}{2} & \frac{1}{2} \\
 \frac{1}{2} & \frac{1}{2} & \frac{1}{2} \\
 \frac{1}{2} & \frac{1}{2} & \frac{1}{2}
\end{array}
\right)
\end{array}
\end{equation*}

and

\begin{equation*}
\kappa_{11kl}(\tcA_{166}^{2})=\left(
\begin{array}{ccc}
 0 & 0 & 0 \\
 0 & 0 & 0 \\
 0 & 0 & 1
\end{array}
\right),\hspace{5mm}
\kappa_{22kl}(\tcA_{166}^{2})=\left(
\begin{array}{ccc}
 0 & 0 & 0 \\
 0 & -1 & 0 \\
 0 & 0 & 1
\end{array}
\right),\hspace{5mm}
\kappa_{33kl}(\tcA_{166}^{2})=\left(
\begin{array}{ccc}
 1 & 1 & 0 \\
 1 & 1 & 0 \\
 0 & 0 & 1
\end{array}
\right)
\end{equation*}

\begin{equation*}
\begin{array}{llllll}
\kappa_{12kl}(\tcA_{166}^{2}) & =\kappa_{21kl}(\tcA_{166}^{2}) & \kappa_{13kl}(\tcA_{166}^{2}) & =\kappa_{31kl}(\tcA_{166}^{2}) & \kappa_{32kl}(\tcA_{166}^{2}) & =\kappa_{23kl}(\tcA_{166}^{2})
\\\\
& =\left(
\begin{array}{ccc}
 0 & 0 & 0 \\
 0 & 0 & 0 \\
 0 & 0 & 1
\end{array}
\right), &
&=\left(
\begin{array}{ccc}
 0 & 0 & 1 \\
 0 & 0 & 1 \\
 1 & 1 & 0
\end{array}
\right), &
& =\left(
\begin{array}{ccc}
 0 & 0 & 1 \\
 0 & 0 & 1 \\
 1 & 1 & 0
\end{array}
\right)
\end{array}
\end{equation*}

The single triangulation comprising geometry 167 gives rise to

\begin{equation*}
\kappa_{11kl}(\tcA_{166}^{2})=\left(
\begin{array}{ccc}
 0 & 0 & 0 \\
 0 & 0 & 0 \\
 0 & 0 & 1
\end{array}
\right),\hspace{5mm}
\kappa_{22kl}(\tcA_{166}^{2})=\left(
\begin{array}{ccc}
 0 & 0 & 0 \\
 0 & 0 & 0 \\
 0 & 0 & 0
\end{array}
\right),\hspace{5mm}
\kappa_{33kl}(\tcA_{166}^{2})=\left(
\begin{array}{ccc}
 1 & 1 & 0 \\
 1 & 0 & 2 \\
 0 & 2 & -2
\end{array}
\right)
\end{equation*}

\begin{equation*}
\begin{array}{llllll}
\kappa_{12kl}(\tcA_{166}^{2}) & =\kappa_{21kl}(\tcA_{166}^{2}) & \kappa_{13kl}(\tcA_{166}^{2}) & =\kappa_{31kl}(\tcA_{166}^{2}) & \kappa_{32kl}(\tcA_{166}^{2}) & =\kappa_{23kl}(\tcA_{166}^{2})
\\\\
& =\left(
\begin{array}{ccc}
 0 & 0 & 0 \\
 0 & 0 & 0 \\
 0 & 0 & 1
\end{array}
\right), &
& =\left(
\begin{array}{ccc}
 0 & 0 & 1 \\
 0 & 0 & 1 \\
 1 & 1 & 0
\end{array}
\right), &
& =\left(
\begin{array}{ccc}
 0 & 0 & 1 \\
 0 & 0 & 0 \\
 1 & 0 & 2
\end{array}
\right)
\end{array}
\end{equation*}

\subsubsection{\textbf{Mori Cone Phase Matrix}}\label{sec:MORIMATN}

\begin{figure}[!h]
\centering
\includegraphics[height=90px]{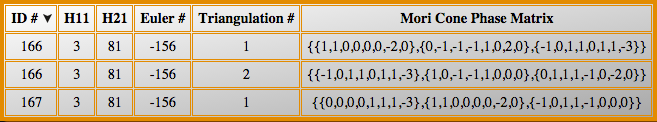}
\label{fig:MORIMATN}
\end{figure}

This is the Mori cone matrix $\mathcal{M}(\tcA^{p})$ for the $p^{\text{th}}$ phase of the desingularized ambient toric variety $\tcA$.\\

See Section \ref{sec:MORIMAT} for more details about the Mori cone and the construction of the Mori cone matrix.\\

In this example, the two triangulations comprising geometry 166 give rise to

\begin{align*}
\mathcal{M}(\tcA_{166}^{1})&=\left(
\begin{array}{cccccccc}
 1 & 1 & 0 & 0 & 0 & 0 & -2 & 0 \\
 0 & -1 & -1 & -1 & 1 & 0 & 2 & 0 \\
 -1 & 0 & 1 & 1 & 0 & 1 & 1 & -3
\end{array}
\right)
\\
\mathcal{M}(\tcA_{166}^{2})&=\left(
\begin{array}{cccccccc}
 -1 & 0 & 1 & 1 & 0 & 1 & 1 & -3 \\
 1 & 0 & -1 & -1 & 1 & 0 & 0 & 0 \\
 0 & 1 & 1 & 1 & -1 & 0 & -2 & 0
\end{array}
\right)
\end{align*}

The single triangulation comprising geometry 167 gives rise to

\begin{equation*}
\mathcal{M}(\tcA_{167})=\left(
\begin{array}{cccccccc}
 0 & 0 & 0 & 0 & 1 & 1 & 1 & -3 \\
 1 & 1 & 0 & 0 & 0 & 0 & -2 & 0 \\
 -1 & 0 & 1 & 1 & -1 & 0 & 0 & 0
\end{array}
\right)
\end{equation*}

Note that because geometry 167 is only composed of a single triangulation, its Mori cone matrix is equivalent to that of the full geometry in Section \ref{sec:MORIMAT}.

\subsubsection{\textbf{K\"ahler Cone Phase Matrix}}\label{sec:KAHLERMATN}

\begin{figure}[!h]
\centering
\includegraphics[height=90px]{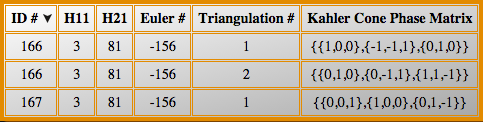}
\label{fig:KAHLERMATN}
\end{figure}

This is the K\"ahler cone matrix $\mathcal{K}^{i}_{j}(\tcA^{p})$ for the $p^{\text{th}}$ phase of the desingularized ambient toric variety $\tcA$.\\

See Section \ref{sec:KAHLERMAT} for more details about the K\"ahler cone and the construction of the K\"ahler cone matrix.
Recall that for favorable geometries, the K\"ahler cone of each phase is simplicial, and therefore these matrices should all be square $h^{1,1}\times h^{1,1}$ matrices.
Furthermore, because the \textbf{Divisor Class Basis} has real coefficients, these matrices will in general be non-trivial (i.e., not equal to the identity matrix).\\

In this example, the two triangulations comprising geometry 166 give rise to

\begin{align*}
\mathcal{K}(\tcA_{166}^{1})&=\left(
\begin{array}{ccc}
 1 & 0 & 0 \\
 -1 & -1 & 1 \\
 0 & 1 & 0
\end{array}
\right)
\\
\mathcal{K}(\tcA_{166}^{2})&=\left(
\begin{array}{ccc}
 0 & 1 & 0 \\
 0 & -1 & 1 \\
 1 & 1 & -1
\end{array}
\right)
\end{align*}

The single triangulation comprising geometry 167 gives rise to

\begin{equation*}
\mathcal{K}(\tcA_{167})=\left(
\begin{array}{ccc}
 0 & 0 & 1 \\
 1 & 0 & 0 \\
 0 & 1 & -1
\end{array}
\right)
\end{equation*}

Note that because geometry 167 is only composed of a single triangulation, its K\"ahler cone matrix is equivalent to that of the full geometry in Section \ref{sec:KAHLERMAT}.

\section{Discussion}\label{sec:conc}

The work described in this paper was motivated primarily by the desire of string theorists for large, easy-to-access, datasets of topological and geometrical properties of Calabi--Yau threefolds.
The traditional tool for approaching the largest such set of Calabi--Yau manifolds, the Kreuzer--Skarke database \cite{Kreuzera}, is \texttt{PALP} (Package for Analyzing Lattice Polytopes).
\texttt{PALP} is a wonderful resource, but it only goes so far in providing physicists with the information they need.

As an example of this, \texttt{PALP} is unable to compute a secondary polytope that is more than three-dimensional.
That is, the triangulation algorithm is coded in such a way as to be limited to polytopes with no more than three points interior to the facets of $\Delta^{*}$.
Therefore, \texttt{PALP}'s ability to triangulate the Kreuzer--Skarke dataset is limited to only the smaller reflexive polytopes.
For example, 377 of the 4990 reflexive polytopes with $h^{1,1} = 5$, or 7.6\%, could not be triangulated with \texttt{PALP}.
At $h^{1,1} = 6$ the fraction of polytopes which could not be triangulated grew to 23.4\% (4007 out of 17,101).
No polytopes with $h^{1,1} > 6$ which can be triangulated using \texttt{PALP} have been identified.

In addition, and perhaps more importantly, deriving all of the data that a physicist may want from the Kreuzer--Skarke list is computationally intensive.
It is a waste of resources for every group interested in such topics to be forced to recompute these results in isolation.
Thus, we present our own dataset that we have compiled from the Kreuzer--Skarke database in the online repository located at \url{http://nuweb1.neu.edu/cydatabase}.

The methods described in Section~\ref{sec:pedagogy} were applied to the 23,568 polytopes with $h^{1,1} \leq 6$.
The number of independent triangulations and glued Calabi--Yau geometries obtained for each value of $h^{1,1}$ is collected in Table~\ref{tab:geometries}.

\vspace{0.2cm}

\begin{table}[!h]
\textbf{\refstepcounter{table}\label{tab:geometries} Table \arabic{table}.} {\bf  Count of Polytopes and Geometries}

\vspace{0.2cm}

{\begin{tabular}{|l|c|c|c|c|c|c|c|}
\hline
$h^{1,1}(X)$ & 2 & 3 & 4 & 5 & 6 & 7 & 8 \\ \hline
Number of Polytopes & 36 & 244 & 1197 & 4990 & 17101 & 50376 & 128165 \\ \hline
Number of Triangulations & 48 & 526 & 5348 & 57050 & 589025$^*$ & -- & -- \\ \hline
Number of Geometries & 39 & 306 & 2014 & 13635 & 85679$^*$ & -- & -- \\ \hline
Number of Favorable Geometries & 39 & 305 & 2000 & 13494 & 84522$^*$ & -- & -- \\ \hline
\end{tabular}}
\caption*{\begin{footnotesize} $^*$Of the 17,101 reflexive polytopes with $h^{1,1}=6$, we find three cases which each fail to triangulate after 2,160 core-hours of processor time. The numbers in this table reflect the results from the remaining 17,098 polytopes.\end{footnotesize}}
\end{table}

\vspace{0.2cm}

The total number of triangulations performed was just under 652,000, resulting in 101,673 CY threefolds.
Of these, more than 100,000 are favorable cases, and are thus amenable to meaningful phenomenological study. Where results could be obtained
Where results could be obtained using \texttt{PALP 2.1}, we find agreement between the \texttt{Sage} implementation and the output from \texttt{PALP}.

\vspace{0.2cm}

We note that while the number of polytopes grows polynomially with $h^{1,1}$, the number of core-hours required to obtain the ultimate Calabi--Yau data grows exponentially.
A (perhaps naive, but illustrative) fit to the computer time needed for $h^{1,1} \leq 5$ suggests that the processor computation time spent obeys roughly
\begin{equation} t = (5\times 10^{-5}) \times e^{3.5\,h^{1,1}}\,{\textnormal{ core-hrs}} \label{corefit} \end{equation}
For perspective, the extraction of the full set of 85,679~geometries at $h^{1,1}=6$ took over 113,430 core-hours to complete.
This is a little less than 80~core-minutes per unique geometry obtained.
To a first approximation, the bulk of the processor time is spent at the triangulation stage, whose computational intensity also grows exponentially.

Work is currently underway to extend these results to $h^{1,1}=7$ and beyond.
If the empirical formula in equation~(\ref{corefit}) continues to hold, this addition will require over 2.2~million core-hours to fully explore.
Clearly, extending the database will require mitigating this additional computational load by using the technique of triangulating maximal cones, as described in Section~\ref{sec:MPCP}.
As the database expands, the newly-computed Calabi--Yau data will be appended to that already hosted on the web-based repository~\cite{Altmana}.\\

\section*{Acknowledgements}
We are grateful to Lara Anderson, Per Berglund, Volker Braun, Stefan Groot Nibbelink, Benjamin Jurke, Seung-Joo Lee, Andr\'e Lukas, Herbie Smith, Xin Gao, and Chuang Sun for many helpful discussions during the development of this database.
We especially thank Joan Sim\'on for collaboration on a parallel work~\cite{Altman} that applies the database developed in this paper to classify when Calabi--Yau threefolds admit large volume vacua.
JG is supported by NSF PHY-1417316.
YHH is supported by the Science and Technology Facilities Council, UK, for grant ST/J00037X/1, the Chinese Ministry of Education, for a Chang-Jiang Chair Professorship at NanKai University, and the city of Tian-Jin for a Qian-Ren Award.
VJ is supported by the South African Research Chairs Initiative of the Department of Science and Technology and National Research Foundation and thanks McGill University for hospitality.
The work of the authors is funded by the U.S.\ National Science Foundation under the grant CCF-1048082, EAGER: CiC: A String Cartography.

\begin{appendices}
\section{Extended Glossary of Basic Terms} \label{appa}
In the text we give references to the literature for each of the technical steps that we take.
In this extended glossary we will describe some of the basic notions which appear in this subject in an informal manner.
This is intended to provide the physicist who is familiar with differential geometry, but not the details of algebraic geometry, with a rough guide to several of the concepts which are necessary to follow these technical references.
It should be said that in paring down these concepts to provide such a quick introduction to them, we are sacrificing some degree of mathematical rigor.

\vspace{0.5cm}

\noindent {\bf Divisors:}
\vspace{0.2cm}

We will require both of the commonly used descriptions of a divisor on the algebraic variety $\cA$:
\begin{itemize}
\item A {\it Weil divisor} is essentially a formal linear sum of irreducible hypersurfaces within a variety.
\begin{eqnarray} \label{W1}
W=\!\!\!\sum_{\substack{\text{codim}(Y)=1\\ Y\text{ irred.}}} {v_{Y}\cdot Y}\, .
\end{eqnarray}
\item A {\it Cartier divisor} is a description of a divisor in terms of a collection of rational functions associated to each coordinate patch on a variety: $(U_i , f_i)$ where $i$ runs over the full open cover of $\cA$.
On the overlap $U_i \cap U_j$ (for any $i$ and $j$) it is required that the transition function $f_i/f_j$ is a non-zero rational function.
We denote the abelian group of Cartier divisors $\mathcal{C}(\cA)$.
\end{itemize}
On smooth varieties, it turns out that the notions of Cartier and Weil divisors coincide.
However, since our work will sometimes involve singular spaces we need to keep the two concepts distinct.

\vspace{0.2cm}

\noindent We shall also need some other notions associated to the concept of divisors:
\begin{itemize}
\item A divisor is said to be {\it effective} if all of the coefficients $v_{Y}$ in (\ref{W1}) are non-negative, or, in the Cartier case, if the $f_i$ can be chosen to be regular functions.
\item A Cartier divisor is said to be {\it principal} if it is described by a globally-defined rational function.
\item Two Cartier divisors $D_1$ and $D_2$ are said to be {\it linearly equivalent} (or more generally, rationally equivalent) if $D_1=D_2+D_p$ where $D_p$ is any principal divisor.
A complete set of divisors linearly equivalent to $D$ is called a {\it divisor class} and is denoted\footnote{Occasionally in the literature, the set of divisors linearly equivalent to $D$ is called a \textit{linear system} and is denoted $|D|$.} $\LD{D}$.
\item The intersection of divisors in $\LD{D}$ is called its \textit{base locus}.
When the base locus is the empty set, we say that the divisor class is \textit{base point free}.
Bertini's theorem tells us that a divisor $D$ is smooth away from its base locus, and it is smooth everywhere when $\LD{D}$ is base point free.
\end{itemize}

\vspace{0.5cm}

\noindent {\bf Chow Ring:}
\vspace{0.2cm}

One can define higher codimension generalizations of divisors called algebraic cycles.
Just like divisors, there is a notion of rational equivalence of algebraic cycles (a generalization of linear equivalence of divisors), which allows for a concept of cycle classes.
We denote the Abelian group of equivalence classes of codimension $k$ cycles on an algebraic variety ${\cal A}$ by $A^k({\cal A})$.
The Chow ring is then defined by
\begin{eqnarray}
A({\cal A}) = \bigoplus_{k\geq 0} A^k({\cal A}) \;.
\end{eqnarray}
In particular, the multiplication structure which endows $A({\cal A})$ with the structure of a ring is  given by intersection of algebraic cycles.
If we have two rational equivalence classes\footnote{A rational equivalence class is a generalization of a linear equivalence class for cycles of arbitrary codimension.} $\RD{C_1}$ and $\RD{C_2}$ in $A^i({\cal A})$ and $A^j({\cal A})$ respectively, then we define the product to be $\RD{C_1} \cdot \RD{C_2} = \RD{C_1 \cap C_2}$, which is an element of $A^{i+j}({\cal A})$.

\vspace{0.5cm}

\noindent {\bf Line Bundles from Divisors:}
\vspace{0.2cm}

Associated to any Cartier divisor $D$ of a variety ${\cal A}$ is a holomorphic rank one vector bundle, or \textit{line bundle}, denoted as ${\cal O}_{\cA}(D)$ over ${\cal A}$.
The ratios $f_i/f_j$ which appear in the description of a Cartier divisor above give the transition functions defining ${\cal O}_{\cA}(D)$.

The Picard group $\text{Pic}(\cA)$ is the abelian group of isomorphism classes of holomorphic line bundles on $\cA$. There exists a relationship between line bundles and Cartier divisors given by $\text{Pic}(\cA)\cong\mathcal{C}/\sim_{lin}$.

\begin{itemize}
\item The first Chern class, we write
\begin{equation*}
c_1({\cal O}_{\cA}(D))= \gamma(D)
\end{equation*}
\noindent where $\gamma$ represents the operation of Poincar\'e duality: taking the cohomology class of the $(1,1)$-form dual to the divisor class of $D$.
\item If the Cartier divisor $D$ corresponds to a subvariety inside ${\cal A}$, then ${\cal O}_{\cA}(D)$ is referred to as the \textit{normal bundle} to $D$ in ${\cal A}$.
\end{itemize}

A line bundle is said to be \textit{very ample} when there are enough global sections to set up an embedding into projective space.
Such a globally-generated line bundle always exists in a projective variety.
A line bundle is said to be \textit{ample} when some positive power is very ample.
By a common abuse of terminology, we sometimes say that the divisor $D$ defining an ample line bundle ${\cal O}_{\cA}(D)$ is ample as well.

Because global sections must be defined everywhere, they are holomorphic.
Therefore, ample Cartier divisors are frequently effective as well.

\vspace{0.5cm}

\noindent {\bf Adjunction:}
\vspace{0.2cm}

Given a divisor $D$ defining a hypersurface in a variety ${\cal A}$ there exists a short exact sequence
\begin{eqnarray}
0 \to TD \to T{\cal A}|_D \to {\cal O}_{\cA}(D)|_D \to0 \;.
\end{eqnarray}
This essentially says that the tangent directions to the manifold ${\cal A}$ at a point on $D$ are those directions tangent to $D$ (encapsulated by $TD$) and those directions normal to it in ${\cal A}$ (encapsulated by the normal bundle ${\cal O}_{\cA}(D)|_D$.
\begin{itemize}
\item Chern classes behave in a particular way under such a short exact sequence.
Namely
\begin{eqnarray}
c(T{\cal A}|_D) = c(TD) \wedge c ({\cal O}_{\cA}(D)|_D)\;.
\end{eqnarray}
In particular, therefore, $c_1(TD) = c_1(T{\cal A}|_D) -  c_1 ({\cal O}_{\cA}(D)|_D) = c_1(T{\cal A}|_D) - \gamma(D)$.
\end{itemize}

\vspace{0.5cm}

\noindent {\bf Calabi--Yau Manifolds as Hypersurfaces in Fano Varieties:}
\vspace{0.2cm}

A Calabi--Yau threefold is a six dimensional K\"ahler manifold with vanishing first Chern class.
We can construct Calabi--Yau manifolds as hypersurfaces $D$ inside an ambient space ${\cal A}$ if $c_1(T{\cal A}|_D) =  c_1 ({\cal O}_{\cA}(D)|_D) = \gamma(D)$.
In particular, we can take $D$ to be a so-called \textit{anticanonical divisor} $-K_{\cal A}$.
\begin{itemize}
\item From the tangent bundle $T{\cal A}$ we can construct its top wedge power ${\cal K}^{\vee}_{\cal A} = \wedge^{\textnormal{dim} {\cal A}} T{\cal A}$ which is a line bundle known as the anticanonical bundle.
\item By the properties of Chern classes it can be shown that $c_1(T{\cal A}) = c_1 ( \wedge^{\textnormal{dim} {\cal A}} T{\cal A}) = c_1 ({\cal K}^{\vee}_{\cal A})$.
\item A Cartier divisor $-K_{\cal A}$ in the class associated to the anticanonical bundle ${\cal K}^{\vee}_{\cal A}$ is called an anticanonical divisor.
In the case where such a divisor defines a codimension one subvariety inside ${\cal A}$ we see by adjunction and the above bullet point that this hypersurface has vanishing first Chern class and is thus Calabi--Yau.
\end{itemize}
The conclusion we reach is that in cases where an anticanonical divisor of a variety ${\cal A}$ defines a subvariety of ${\cal A}$, that subvariety is a Calabi--Yau manifold.
In other words, $X=-K_{\cA}$ is a Calabi--Yau manifold if $-K_{\cA}$ is a subvariety of $\cA$.
This is always the case when $-K_{\cA}$ is ample, which occurs by definition when $\cA$ is a \textit{Fano variety}.

\vspace{0.5cm}

\noindent {\bf Toric Varieties:}
\vspace{0.2cm}

A toric variety $\cA$ is defined as an algebraic variety containing a torus $T$ as a dense open subset such that the action of $T$ on itself extends to all of $\cA$, i.e., $T\times\cA\rightarrow\cA$.

An $n$-dimensional toric variety can be described as a quotient
\begin{equation}\label{eq:toric1}
\cA\cong \frac{V}{(C^{*})^{k-n}\times G}\, ,
\end{equation}
where $V$ is $\mathbb{C}^k$ with some ``exceptional set'' excised and $G$ is the group of orbifold automorphisms taking $\cA$ to itself.
In many applications in physics the group $G$ can be taken to be trivial and we are left with the simple split torus action $T=(C^{*})^{k-n}$ in the denominator of the quotient.

In a toric variety, the zero locus of each coordinate of $V$ can be associated to a \textit{toric divisor class} $z_{i}\mapsto D_{i}=\{z_{i}=0\}$, however this map does not take into account linear equivalence. Therefore, the Picard group of holomorphic line bundle classes over $\cA$ is lower dimensional. It can be shown that for a toric variety, $\text{Pic}(\cA)\cong\mathbb{Z}^{k-n}$.   

It is important to note that toric varieties are often singular.
This will play a key role in the discussion in the text as we will frequently start with a singular variety and then (partially) resolve it in several different ways.

\vspace{0.5cm}

\noindent {\bf Gorenstein Toric Fano Varieties and Reflexive Polytopes:}
\vspace{0.2cm}

A toric Fano variety is a toric variety with an anticanonical divisor which is an ample Cartier divisor.
From the discussion above, this is exactly what we want. In this case, an anticanonical divisor inside a Gorenstein toric Fano variety ${\cal A}$ defines a codimension one subvariety $X\subset{\cal A}$ which is a Calabi--Yau manifold.

In order for a construction of Calabi--Yau manifolds to have substantial computational power, we must be able to determine the relevant quantities of these manifolds using combinatorics and linear algebra.
Happily, this is possible in the Gorenstein toric case due to the one-to-one correspondence between Gorenstein toric Fano varieties and reflexive polytopes.
\begin{itemize}
\item A lattice polytope $\Delta$ is the convex hull of finitely many vertices on an integer lattice $M\cong\mathbb{Z}^{n}$.
\item We then define a dual lattice $N\cong\mathbb{Z}^{n}$ parameterized by the maps taking the $M$ into the integers, i.e., $N=\textnormal{Hom}_{\mathbb{Z}} (M, \mathbb{Z})$.
\item Then, on $N$, we can define the dual (or polar) polytope by
\begin{equation}\label{eq:dualpoly}
\Delta^*=\{\bm{n} \in N\; |\;\langle\bm{m},\bm{n}\rangle\geq -1, \;\forall\bm{m}\in\Delta \}\; .
\end{equation}
\item A lattice polytope $\Delta\subset M$ containing only the origin of $M$ in its interior is said to be \textit{reflexive} if $\Delta^*\subset N$ is also a lattice polytope containing only the origin of $N$ in its interior.
\item Such reflexive polytopes are in one-to-one correspondence (up to birational equivalence) with Gorenstein toric Fano varieties.
In particular, we can derive many properties of such varieties from simple combinatorial operations involving the polytope.
\end{itemize}

Below we give an example of how information about a Gorenstein toric Fano variety ${\cal A}$ can be extracted from the data of a polytope.
In particular, we will briefly describe how one obtains some information about divisors in ${\cal A}$.

\vspace{0.5cm}

\noindent {\bf Divisors From Polytopes:}
\vspace{0.2cm}

A heuristic description of how one extracts information about divisors given a polytope $\Delta^*$ is as follows
\begin{itemize}
\item For each point $\bm{m}\in\Delta$, define the \textit{supporting hyperplane} $H_{\bm{m}}=\{\bm{n}\in N\; |\;\langle\bm{m},\bm{n}\rangle=-1\}$.
Then, by equation (\ref{eq:dualpoly}), we see that $\Delta^{*}$ is bounded by the hyperplanes $H_{\bm{m}}$.
\item Then, a \textit{facet}, or ($n-1$)-dimensional face of $\Delta^{*}$, is given by $F=\Delta^{*}\cap H$ for $H$ a supporting hyperplane.
We define the set $\mathcal{F}$ of facets and all their intersections.
Then, the subset of ($d-1$)-dimensional faces is denoted $\mathcal{F}_{d-1}$.
\item We define the ($d-1$)-skeleton $\text{skel}_{d-1}$ to be the union of faces of dimension $\leq d-1$.
\item A $d$-dimensional convex, rational, polyhedral \textit{cone} is given by $\sigma=\text{cone}(F)$ for $F\in\mathcal{F}_{d-1}$, where $\text{cone}(F)$ is the set of all rays that pass from the origin through points in $F$.
The $n$-dimensional cones are called \textit{maximal cones}, and we define the set $\Sigma$, called the \textit{fan} of $\Delta^{*}$, of maximal cones and all their intersections.
Then, the subset of $d$-dimensional cones is denoted $\Sigma_{d}$, and their union $|\Sigma_{d}|=\bigcup_{\sigma\in\Sigma_{d}}{\sigma}$ is called the $d$\textit{-support} of $\Delta^{*}$.
\item One-dimensional rays (or more precisely primitive generators of $\Sigma_{1}$) in this fan correspond one-to-one with divisors on the Gorenstein toric Fano variety $\cA$.
These are just the toric divisor classes, denoted by $D_i$.
\item A Calabi--Yau hypersurface, defined by the anticanonical divisor of the Gorenstein toric Fano variety, can then be written in terms of the toric divisor classes
\begin{eqnarray}
X=-K_{\cal A} = \sum_{i}{D_i} ~.
\end{eqnarray}
\item A generic Calabi--Yau hypersurface can also be written in terms of the vanishing of a homogenous Laurent polynomial $P_{X}(\bm{z})$ by
\begin{equation}
P_{X}(\bm{z})=\sum_{\bm{m}\in\Delta\cap M}{c_{\bm{m}}\prod_{i}{z_{i}^{\langle\bm{m},\bm{n}_{i}\rangle +1}}} ~,
\end{equation}
where $c_{\bm{m}}$ are arbitrary coefficients, the choice of which is related to the complex structure on $X$.
Because the exponents of the monomial terms are related to the points of the polytope $\Delta$, we sometimes refer to it as a \textit{Newton polytope}.
\end{itemize}

\section{Nomenclature}
\label{glossary}

Some of the mathematical background which is required in this work is reviewed in simple terms in Appendix~\ref{appa}.
Here, we define some of the notation which appears in the main discussion of this paper.

\begin{itemize}
\item
$\cA$ is an ambient Gorenstein toric Fano variety.
\item $\Delta$ is a reflexive Newton polytope corresponding to the ambient variety $\cA$.
\item The desingularization of $\cA$ is $\tcA$.
\item $\mathcal{V}(\Delta^{*})$ is the set of vertices of the reflexive polytope $\Delta^{*}$ before subdividing.
\item $\mathcal{P}$ is the set of vertices of $\Delta^{*}$ after subdividing.
\item $M\cong\text{Hom}(T,\mathbb{C}^{*})\cong\mathbb{Z}^{n}$ is the character lattice of the split torus $T=(\mathbb{C}^{*})^{k-n}$.
\item $N\cong\text{Hom}(M,\mathbb{Z})\cong\mathbb{Z}^{n}$ is the dual lattice.
\item $\langle ,\rangle:\; M\times N\rightarrow\mathbb{Z}$ is the inner product between dual lattices.
\item $\text{card}(P)$ is the cardinality, or number of lattice points, in any subspace $P\subset M$ or $N$.
\item In this paper, we use $n=\text{dim}(\cA)=\text{dim}(\tcA)$, $k=\text{card}(\mathcal{P})$, and $d$ an arbitrary dimension.
\item $\text{relint}(P)$ is the relative interior of any subspace $P\subset M$ or $N$ and $\partial P$ is its boundary.
\item $[P]$ is the $n\times\text{card}(P)$ matrix with columns given by the vectors $\bm{p}\in P$ for $P$ some point configuration.
\item
Occasionally, for the sake of organization and reproducibility, we will use $\text{sort}(P)$ to signify the poset of lattice points in a subspace $P\subset M$ or $N$, sorted in ascending order by their coordinate values in $M$ or $N$.
\item $H_{\bm{m}}(\Delta^{*})=\{\bm{n}\in N\;\vert\;\langle\bm{m},\bm{n}\rangle=-1\}$ is a supporting hyperplane of $\Delta^{*}$ corresponding to the point $\bm{m}\in\Delta$.
\item $\Delta^{*}=\{\bm{n}\in N\;\vert\;\langle\bm{m},\bm{n}\rangle\geq -1,\;\forall\bm{m}\in\Delta\}$ is the dual (or polar) polytope, which is also equal to the intersection of half-spaces bounded by the $H_{\bm{m}}(\Delta^{*}),\;\bm{m}\in\Delta$.
\item The faces $F\in\mathcal{F}(\Delta^{*})$ of $\Delta^{*}$ are formed by all possible intersections of the supporting hyperplanes $H_{\bm{m}}(\Delta^{*}),\;\bm{m}\in\Delta$ with $\Delta^{*}$.
The subset of $(d-1)$-faces $\mathcal{F}_{d-1}(\Delta^{*})\subset\mathcal{F}(\Delta^{*})$ is the set of $(d-1)$-dimensional faces.
\item The ($d-1$)-skeleton $\text{skel}_{d-1}$ is the union of faces $F$ of dimension $\leq d-1$.
We see that $\mathcal{V}(\Delta^{*})=\text{skel}_{0}(\Delta^{*})$.
\item For each face $F\in\mathcal{F}(\Delta^{*})$, there is a corresponding dual face $F^{*}\in\mathcal{F}(\Delta)$ given by $F^{*}=\{\bm{m}\in\Delta\;\vert\;H_{\bm{m}}(\Delta^{*})\cap F\neq\emptyset\}$.
\item For every face $F\in\mathcal{F}(\Delta^{*})$, there is a corresponding convex rational polyhedral cone $\sigma_{F}=\text{cone}(F)$ formed by the space of rays from the origin passing through $F$.
\item
$\text{rays}(\sigma)$ is the set of extremal rays of the convex, polyhedral cone $\sigma$.
Note that $\text{cone}(\text{rays}(\sigma))=\sigma$.
\item $\Sigma(\Delta^{*})=\{\sigma_{F}\;\vert\; F\in\mathcal{F}(\Delta^{*})\}$ is the fan of $\Delta^{*}$.
The subset of $d$-cones is $\Sigma_{d}(\Delta^{*})=\{\sigma_{F}\;\vert\; F\in\mathcal{F}_{d-1}(\Delta^{*})\}$.
\item For each cone $\sigma\in\Sigma(\Delta^{*})$, there is a corresponding dual cone $\sigma^{*}\in\Sigma(\Delta)$ given by $\sigma^{*}=\{\bm{m}\in M\;\vert\;\langle\bm{m},\bm{n}\rangle\geq 0,\;\bm{n}\in\sigma\}$.
\item $T(\tcA)$ is the triangulation of $\Delta^{*}$ corresponding to the desingularization $\tcA$ of $\cA$.
The set of all such triangulations is $\mathcal{T}$.
\item
$\text{vol}(S)$ is the lattice volume of a simplex $S\in T(\tcA)$, defined to be the geometric, oriented volume of $S$ divided by the volume 1/n! of a unit simplex~\cite{Knapp:2011ip}.
\item $\mathcal{U}(\tcA)=\left\{U\subset\tcA\;\left\lvert\;\tcA\cong\bigcup{U}\right.\right\}$ is an open cover of $\tcA$.
\item $D_{1},...,D_{k}$ are toric divisor classes on $\tcA$, and $J_{1},...,J_{k-n}$ are the basis elements of the space of divisor classes on $\tcA$.
\item Vertices $\bm{m}_{U}\in\mathcal{V}(\Delta)$, facets $F_{U}\in\mathcal{F}_{n-1}(\Delta^{*})$, and maximal cones in $\sigma_{U}\in\Sigma(\Delta^{*})$ are each in one-to-one correspondence with coordinate patches $U\in\mathcal{U}(\tcA)$ with bijections $\bm{m}_{U}\rightarrow U$, $F_{U}\rightarrow U$, and $\sigma_{U}\rightarrow U$.
\end{itemize}

\end{appendices}





\end{document}